\documentclass[12pt]{article}
\usepackage{amsmath,amssymb,amsfonts,amsthm,bbm}
\usepackage{epic,eepic,epsfig,longtable}
\usepackage{multirow,verbatim}
\usepackage{array}
\usepackage{graphicx}
\usepackage{subcaption}
\usepackage{floatrow}
\usepackage{paralist}
\usepackage{latexsym}
\usepackage{comment}
\usepackage{epsfig}
\usepackage{setspace}
\usepackage{CJK}
\usepackage{color, soul}
\RequirePackage[colorlinks,citecolor=blue,urlcolor=blue]{hyperref}
\usepackage{rotating}

\usepackage[authoryear,round]{natbib}
\makeatletter
\def\singlespace{\def\baselinestretch{1}\@normalsize}

\makeatletter
\def\singlespace{\def\baselinestretch{1}\@normalsize}

\numberwithin{equation}{section}

\renewcommand{\hat}{\widehat}

\renewcommand{\hat}{\widehat}

\newcommand{\bfm}[1]{\ensuremath{\mathbf{#1}}}

\def\ba{\bfm a}   \def\bA{\bfm A}  
\def\bb{\bfm b}   \def\bB{\bfm B}  
     
   \def\bD{\bfm D}  
\def\be{\bfm e}   \def\bE{\bfm E}  
\def\bff{\bfm f}  \def\bF{\bfm F}  \def\FF{\mathbb{F}}
     
   \def\bH{\bfm H}  
   \def\bI{\bfm I}

   \def\bL{\bfm L}  
     
     \def\NN{\mathbb{N}}
     
   \def\bP{\bfm P}  
\def\bq{\bfm q}     
     \def\RR{\mathbb{R}}
     \def\SS{\mathbb{S}}
     
\def\bu{\bfm u}   \def\bU{\bfm U}  
   \def\bV{\bfm V}  
\def\bw{\bfm w}   \def\bW{\bfm W}  
   \def\bX{\bfm X}  
\def\by{\bfm y}   \def\bY{\bfm Y}  
   \def\bZ{\bfm Z}

\def\calF{{\cal  F}}

\newcommand{\bfsym}[1]{\ensuremath{\boldsymbol{#1}}}

 \def\balpha{\bfsym \alpha}
 \def\bbeta{\bfsym \beta}               
              \def\bGamma{\bfsym \Gamma}

 \def\bmu{\bfsym {\mu}}                 
 \def\bnu{\bfsym {\nu}}
 \def\btheta{\bfsym {\theta}}           \def\bPsi  {\bfsym {\Theta}}
           
 \def\bsigma{\bfsym \sigma}             \def\bSigma{\bfsym \Sigma}
 \def\blambda {\bfsym {\lambda}}

        \def\bvarrho{\bfsym {\varrho}}
 \def\bPsi{\bfsym {\Psi}}
\def\bvartheta{\bfsym{\vartheta}}	
\def\bTheta{\bfsym{\Theta}}	
 
 \def \bPhi {\bfsym \Phi}

\def\1{\bfsym{1}}	

                 \def\hbbeta{\hat{\bfsym \beta}}
               
              \def\hbGamma{\hat{\bfsym \Gamma}}

              \def\hbtheta {\hat{\bfsym {\theta}}}

 \def\hlambda{\hat{\lambda}}

\DeclareMathOperator{\diag}{diag}
\DeclareMathOperator{\E}{E}

\def\newpage{\vfill\eject}

\def\today{\ifcase\month\or
  January\or February\or March\or April\or May\or June\or
  July\or August\or September\or October\or November\or December\fi
  \space\number\day, \number\year}

\newdimen\biblioindent    \biblioindent=30pt

 at 8truept

\newcommand{\beq}{\begin{equation}}
  \newcommand{\eeq}{\end{equation}}
\newcommand{\beqn}{\begin{eqnarray}}
  \newcommand{\eeqn}{\end{eqnarray}}
\newcommand{\beqnn}{\begin{eqnarray*}}
  \newcommand{\eeqnn}{\end{eqnarray*}}

\allowdisplaybreaks
\usepackage{verbatim}
\def\tilde{\widetilde}

\def\FF{\mathcal{F}}
\def\[{\left [}  \def\]{\right ]} \def\({\left (}  \def\){\right )}
 \def\endpf{$\blacksquare$}
\def\hat{\widehat}

\newtheorem{thm}{Theorem}[section]

\newtheorem{lemma}{Lemma}[section]
\theoremstyle{remark}
\newtheorem{remark}{Remark}
\theoremstyle{proposition}
\newtheorem{proposition}{Proposition}[section]

\newtheorem{assumption}{Assumption}

\def \diag {\mathrm{diag}} \def \Diag {\mathrm{Diag}} 
\def \det {\mathrm{det}} 
\def \Tr { \mathrm{Tr}} 
\def\vech {\mathrm{vech}}

\def \vec{\mathrm{vec}}

\title{Unified Discrete-Time Factor Stochastic Volatility and
Continuous-Time It\^o Models for Combining Inference
Based on Low-Frequency and High-Frequency} 

\author{Donggyu Kim$^1$, Xinyu Song$^2$, and Yazhen Wang$^3$  
\\$^1$ College of Business, \\
Korea Advanced Institute of Science and Technology (KAIST) \\
$^2$ School of Statistics and Management, \\
Shanghai University of Finance and Economics \\
University of Wisconsin-Madison}

\begin{document}
\maketitle

\begin{abstract}
This paper introduces unified models for high-dimensional factor-based It\^o process, 
which can accommodate both continuous-time It\^o diffusion and discrete-time stochastic volatility (SV) models by embedding the discrete SV model in the continuous instantaneous factor volatility process. 
We call it the SV-It\^o model. 
Based on the series of daily integrated factor volatility matrix estimators, we propose quasi-maximum likelihood and least squares estimation methods. 
Their asymptotic properties are established. 
We apply the proposed method to predict future vast volatility matrix whose asymptotic behaviors are studied. 
A simulation study is conducted to check the finite sample performance of the proposed estimation and prediction method. 
An empirical analysis is carried out to demonstrate the advantage of the SV-It\^o model in volatility prediction and portfolio allocation problems. 
\end{abstract}




\newpage
\renewcommand{\baselinestretch}{1.66}
\baselineskip=20pt

\section{Introduction}

Volatility analysis with high-frequency data is a vibrant research area in finance.
Researchers have devoted to developing volatility estimation methods under the continuous-time frameworks such as It\^{o} diffusion process. 
Example estimators for daily volatility include: 
two-time scale realized volatility \citep{zhang2005tale}, 
multi-scale realized volatility \citep{zhang2006efficient, zhang2011estimating},
kernel realized volatility \citep{barndorff2008designing, barndorff2011multivariate},
pre-averaging realized volatility \citep{christensen2010pre, jacod2009microstructure}, 
quasi-maximum likelihood estimator \citep{ait2010high, xiu2010quasi}, 
local method of moments \citep{bibinger2014estimating}, 
and robust pre-averaging realized volatility \citep{fan2018robust}.
These non-parametric estimation methods can estimate historical volatilities very well.
However, in financial practices, we often need to predict future volatilities for the purpose of risk management and portfolio allocation, and these non-parametric methods cannot capture the market dynamics effectively for volatility prediction.
Therefore, we shift our focus to parametric models that are known to be able to account for the financial market dynamics.

Given low-frequency data such as daily log returns, researchers introduced well-performing discrete-time models such as GARCH, stochastic volatility (SV), and vector autoregression (VAR) models to explain market dynamics by using historical volatilities and returns as innovations. 
These models are widely used for empirical financial analyses as they are easy to implement and can explain the market dynamics effectively. 
Thus, it is natural for researchers to harness the discrete-time model structures in the continuous-time frameworks. 
See \citet{engle2006multiple, hansen2012realized, kim2016unified, shephard2010realising} for related research works. 
In their proposed methods, researchers used non-parametric realized volatility estimators from high-frequency data to make inferences for parametric discrete-time models at the low-frequency. 
Their empirical studies demonstrated that for a finite number of assets, by combining the low- and high-frequency methods, the proposed models perform better in volatility prediction tasks than traditional models with low- or high-frequency data alone.
In financial applications, we often encounter a large number of assets, and models designed for the finite dimension become inconsistent and suffer from the curse of dimensionality. 
Thus, in this paper, we explore the approach to unify the discrete-time and continuous-time models appropriately under the high-dimensional set-up for the purpose of vast volatility matrix estimation and prediction.

To overcome the curse of dimensionality, it is common to impose sparsity on the entire vast volatility matrix \citep{bickel2008covariance, cai2011adaptive, kim2016asymptotic, tao2013optimal, wang2010vast}.
However, in finance, there exist common market factors such as industry sectors, inflation reports, Fed rate hikes, and oil prices, which affect the entire market. 
Therefore, assets are widely correlated, and the sparse condition imposed on the entire volatility matrix is not appropriate. 
To model this stylized feature, approximate factor models which indicate that the volatility matrix consists of the low-rank factor volatility and the sparse idiosyncratic volatility matrices were often used \citep{ait2017using, fan2018robust, fan2013large, fan2015incorporating, kim2019factor}. 
Under this set-up, \citet{kim2019factor} further described the eigenvalue process of the latent factor volatility matrix by a unified GARCH-It\^o model structure \citep{kim2016unified}. 
In their proposal, the daily integrated eigenvalues of the latent factor volatility matrix process have historical squared factor returns as the innovations.

In light of the pioneer works, we model the high-frequency data from a large number of assets by the high-dimensional factor-based It\^o process that consists of the latent factor and the idiosyncratic diffusion processes, and further introduce the SV model structure to daily factor integrated volatility matrices that have the finite rank. 
In specific, we develop a continuous instantaneous factor volatility process which has an autoregressive (AR) structure at integer times so that it is a form of some interpolation of the AR structure. 
The instantaneous factor volatility process reflects the current market dynamics while its daily integrated volatility matrices retain the exact AR structure as in the SV model. 
We name our proposal the SV-It\^o model. 
When estimating the factor volatility matrix that is latent, we face an identifiability issue. 
To overcome this, researchers often impose some structures on the latent factor loading and factor volatility matrices such as the orthonormal and diagonal conditions used in \citet{kim2019factor}. 
These conditions are restrictive in the sense that dynamics among market factors cannot be studied.
In this paper, instead of imposing the strong diagonal condition on the factor volatility matrices, we assume stationary condition and follow the factor volatility matrix estimation procedure described in \citet{tao2011large}.
Based on the series of daily factor volatility matrix estimators and the imposed AR structure, we develop quasi-maximum likelihood estimation (QMLE) and least squares estimation (LSE) methods for model parameters whose asymptotic properties are established. 
With the proposed QMLE or LSE, as well as the imposed AR structure, we can estimate future factor volatility matrices effectively. 
On the other hand, the idiosyncratic volatilities are related to firm-specific risk so that the correlations among assets are weak. 
Thus, we impose the common sparse condition on the idiosyncratic volatility matrix and assume it to be time-invariant. 
When estimating the idiosyncratic volatility matrix, we harness its sparse structure and follow the principal orthogonal complement thresholding (POET) estimation procedure proposed by \cite{fan2013large, fan2015incorporating}. 
Combining the future factor and idiosyncratic volatility matrix estimators, we propose an estimator for future vast volatility matrix and examine its asymptotic properties.

This paper is organized as follows.
Section \ref{SEC-2} introduces a unified SV-It\^o model under the high-dimensional factor-based It\^o diffusion process and investigates its properties. 
Section \ref{SEC-3} develops non-parametric estimation methods for the latent factor loading and factor volatility matrices. 
We propose quasi-maximum likelihood estimation and least squares estimation procedures based on the low-rank factor volatilities. 
The asymptotic properties of proposed estimation methods are established. 
Section \ref{SEC-4} proposes an estimator for the conditional expected vast volatility matrix and investigates its asymptotic properties. 
Section \ref{SEC-5} presents numerical illustrations on the proposed methodologies and Section \ref{SEC-6} concludes the paper.
Proofs are collected in Section \ref{SEC-Proof}.


\section{Unified discrete-time and continuous-time models} \label{SEC-2}

\subsection{Discrete-time and continuous-time models}

Both the discrete-time models such as GARCH and SV as well as the continuous-time models such as OU and CIR provide stochastic methods for financial data analyses. 
Discrete-time models are relatively simple parametric models and are often adopted to model the dynamic evolution of the volatility process based on the low-frequency data. 
Continuous-time models are described by more complicated stochastic differential equations instead and can provide non-parametric estimators for daily volatility based on the high-frequency data.
These two types of models have unique characteristics and are not compatible. 
Since the low- and high-frequency data hold trading information for the same asset, it is natural to develop unified models and draw combined inferences. 
Some of the recent attempts include \citet{engle2006multiple, hansen2012realized, kim2016unified, kim2019factor, shephard2010realising, tao2011large}.
In this paper, we introduce unified discrete-time SV and continuous-time factor-based It\^o diffusion models by allowing both the number of low-frequency and high-frequency observations, as well as the number of assets, go to infinity.

\subsection{Notation}

For any given  $p_1$-by-$p_2$ matrix $\bA = \left(A_{ij}\right)$, we denote its spectral norm by $\|\bU\|_2$, its Frobenius norm by $\|\bA\|_F = \sqrt{ \Tr (\bA^{\top} \bA) }$, and its max norm by $\| \bA \| _{\max} = \max_{i,j} | A_{ij}|$.
Moreover, $\vec (\cdot)$ denotes the operator that stacks the columns of a matrix, $\vec^{-1} (\cdot)$ is the inverse operator of $\vec(\cdot)$, and
$\vech(\cdot  )$ is a column vector obtained by vectorizing only the lower triangular part of a matrix.
$\Diag (\cdot)$  returns a square diagonal matrix with the elements of a vector  on the main diagonal, and $\diag (\cdot)$ returns a column vector of the main diagonal elements of a matrix. 
Also let $\det(\bA)$ be the determinant of a matrix $\bA$ and $\Tr(\bA)$ be the trace of $\bA$. 
Let $C$ be a generic constant whose values are free of $n, m,$ and $p$, and may change from appearance to appearance.

\subsection{Unified models}

Let $\bX _t = \(X_{1,t} , \ldots, X_{p,t} \)^{\top}$ be the vector of true underlying log prices of $p$ assets at time $t$. 
In finance, we usually assume that high-frequency data $\bX_t$ obey some continuous diffusion process. 
To account for common market factors in financial industry, we consider the following factor-based diffusion process: 
	\begin{equation} \label{model1}
	d \bX_t = \bmu_t dt + \bL  d \bff_t + d \bu_t,
	\end{equation}
where $\bmu_t \in \RR^{p}$ is a drift vector,  $\bL$ is a $p$-by-$r$ factor loading matrix and $r$ is the total number of market factors.
Moreover, $\bff_t$ and $\bu_t$ are the factor and idiosyncratic diffusion processes, respectively, and obey the following
	\begin{equation*}
	d\bff_t =  \bsigma_t  ^{\top} d \bB_t \qquad \text{and} \qquad d\bu_t =  \bvartheta _t ^{\top} d \bW_t,
	\end{equation*}
where $\bsigma_t$ is an $r$-by-$r$ matrix, $\bvartheta_t$ is a $p$-by-$p$ matrix, $\bB_t$ and $\bW_t$ are independent $r$-dimensional and $p$-dimensional Brownian motions, respectively. 
The stochastic processes $\bmu_t$, $\bsigma_t$, and $\bvartheta_t$ are defined on a filtered probability space $\( \Omega, \calF, \{\calF_t, t\in [0, \infty)\}, P \)$ with filtration $\calF_t$ satisfying the usual conditions.
The daily integrated volatility matrices follow to be 
	\begin{equation*}
	\bGamma_{k} =\bL \bPsi_k \bL ^{\top} + \bGamma_{k} ^s \quad \text{ for } k=1,2, \ldots, n,
	\end{equation*}
where $\bPsi_k =   \int_{k-1}^k \bsigma_t ^{\top} \bsigma_t dt $ and $\bGamma_k ^s = \int_{k-1} ^k \bvartheta_t  ^{\top} \bvartheta_t dt$.

We note that the idiosyncratic process $\bu_t$ corresponds to the firm-specific risk so that its volatility matrix may be sparse. 
Moreover, since firm-specific risk is generally unpredictable, investors seek to minimize its negative impact on an investment portfolio by diversification or hedging.
On the other hand, the latent factor process $\bff_t$ corresponds to the systematic risk or undiversifiable risk that affects the whole market. 
Thus, it is natural to capture market dynamics by modeling the latent factor process. 
In light of this, we propose the following latent factor diffusion process $\bff_t$ that embeds the SV model. 
Define the instantaneous volatility process of $\bff_t$ by
	\begin{equation*}
	\bSigma_t = \bsigma_t ^{\top}  \bsigma_t
	\end{equation*}
and 
	\begin{eqnarray}  \label{VAR}
	\bSigma_t  &=& \bSigma _{[t]} + (t-[t]) \( \balpha_0 \balpha_0 ^{\top} - \bSigma _{[t]}  + \sum_{j=1}^{q-1} \balpha_{j+1} \bPsi _{[t]-j+1} \balpha_{j+1} ^{\top} \)  \cr
	&&+ \balpha_{1} \(\int_{[t]} ^{t} \bSigma_s ds \) \balpha_1 ^{\top} + ([t] +1 -t) \bZ _t \bZ_t^{\top},
	\end{eqnarray}
where $\bZ_t = (Z_{i,t}) _{i=1,\ldots,r} = \int_{[t]} ^{t} \bnu ^{\top} d \bB_s ^1$, $\bB_t^1$ is an $r$-dimensional standard Brownian motion, 
$[t]$ denotes the integer part of $t$ except that $[t] = t -1$ when $t$ is an integer,   
and $\balpha_j$'s are $r$-by-$r$ matrices. 
We name our proposal the SV-It\^{o} model.

The instantaneous volatility process $\bSigma_t$ is almost surely continuous with respect to time $t$. 
When restricted the integer time points, $\bSigma_t$ of the SV-It\^o model retains the following AR structure,  
	\begin{equation*}
	\bSigma_k  = \balpha_0 \balpha_0 ^{\top} + \sum_{j=1}^q \balpha_{j} \bPsi _{k-j+1} \balpha_j ^{\top} \quad  \text{for } k=1, 2,\ldots,n.  
	\end{equation*}
Thus, the instantaneous volatility process is formed by some interpolation of the AR structure where the random fluctuation is explained by the $\bZ_t$ process. 
Also, current market dynamics are reflected through the terms  $\int_{[t]} ^{t} \bSigma_s ds$ and $\bZ_t$.

For statistical inferences, we have an interest in developing parametric models based on the series of daily integrated factor volatility matrices, i.e., $\int_{k-1}^k \bSigma_t dt$, $k=1,\ldots,n$. 
In the following proposition, we show that the daily integrated factor volatility matrices have the discrete-time AR structure.

\begin{proposition}
\label{Prop-VAR}
Under the SV-It\^o model, we have the following iterative relations.
\begin{enumerate}
\item [(a)] For any $n, k \in \NN$, we have 
	\begin{eqnarray*}
	&&\bF(k)  \equiv  \int_{n-1}^n \frac{(n-t)^k}{k!} \vec( \bSigma_t ) dt   \cr
	&&=   \frac{ \vec (\balpha_0 \balpha_0 ^{\top})  + (k+1)  \vec( \bSigma _{n-1}  )  + \sum_{j=1}^{q-1} \bA_{j+1}  \vec(\bPsi _{n-j } )  }{(k+2)!}  + \frac{k+1}{(k+3)!}  \vec(\bnu ^{\top} \bnu ) \cr
	&&   + \vec \(  \( \int_{n-1}^n \frac{(k+1) (n-t) ^{k+2}}{ (k+2)!} Z_{j,t} d Z_{i,t} + \int_{n-1}^n \frac{(k+1) (n-t) ^{k+2}}{ (k+2)!} Z_{i,t} d Z_{j,t} \) _{i,j=1,\ldots, r} \) \cr
	&&  + \bA_{1} \bF(k+1),
	\end{eqnarray*}
where $\bA_j =\balpha_{j} \otimes \balpha_{j} $ for $j=1,\ldots,q$, and the operator $\otimes$ denotes the Kronecker product.

\item[(b)]
For $\det(\balpha_1) \neq 0$ and $\|\balpha_1\|_2 <1$, we have
	\begin{eqnarray*}
	\int_{n-1}^n  \vec( \bSigma_t ) dt  &=&   \bvarrho_1  \vec (\balpha_0 \balpha_0 ^{\top} ) + \( \bvarrho_2 -  2 \bvarrho_3 \) \vec(\bnu ^{\top} \bnu ) \cr
	&&+ \sum_{j=1}^{q-1} \{  (\bvarrho_1-   \bvarrho_2) \bA_{j } +   \bvarrho_2 \bA_{j+1}  \} \vec(\bPsi _{n-j} )  \cr
	&&+ (\bvarrho_1-   \bvarrho_2) \bA_{q }  \vec(\bPsi _{n-q} ) + \bD_n \text{ a.s.},
	\end{eqnarray*} 
where
$\bvarrho_1= \bA_1 ^{-1} \( e ^{\bA_1} - \bI_{r^2}  \)$, 
$\bvarrho_2= \bA_1 ^{-2} \( e ^{\bA_1} - \bI_{r^2} - \bA_1 \)$, 
$\bvarrho_3= \bA_1 ^{-3} ( e ^{\bA_1} - \bI_{r^2} - \bA_1 - \frac{\bA_1 ^2}{2})$, 
$\bI_{r^2}$ is the $r^2$-dimensional identity matrix, $e^{\bA} = \sum_{k=0}^\infty \frac{\bA^k} {k!}$, 
and
	\begin{eqnarray*}
	\bD_n &=& \sum_{k=0}^{\infty}  \bA_1 ^k \vec \Big  ( \Big ( \int_{n-1}^n \frac{(k+1) (n-t) ^{k+2}}{ (k+2)!} Z_{j,t} d Z_{i,t} \cr
	&& \qquad \qquad \qquad \qquad \qquad + \int_{n-1}^n \frac{(k+1) (n-t) ^{k+2}}{ (k+2)!} Z_{i,t} d Z_{j,t} \Big ) _{i,j=1,\ldots, r} \Big ).
	\end{eqnarray*}
\end{enumerate}
\end{proposition}

Proposition \ref{Prop-VAR} shows that the daily integrated volatility matrices retain some AR structure. 
Moreover, by the construction, there exist $r(r+1)/2$ vector $\bbeta_0$ and $r(r+1)/2$-by-$r(r+1)/2$ matrices $\bbeta_j, j=1,\ldots, q$, such that 
	\begin{equation} \label{VAR-int}
	\vech \( \bPsi _{n} \) = \bbeta_0 + \sum_{j=1}^q \bbeta_j \vech \( \bPsi _{n-j} \) +\be_n ,
	\end{equation}
where $ \bbeta_0 =\vech \(  \vec ^{-1} \( \bvarrho_1  \vec (\balpha_0 \balpha_0 ^{\top} ) + \( \bvarrho_2 -  2 \bvarrho_3 \) \vec(\bnu ^{\top} \bnu ) \) \)$, 
and the $i$th row of $\bbeta_j$ is obtained by $ \vech  ( \vec ^{-1} \( \bA \) +   \vec ^{-1} \( \bA ^{\top}  \)  - \Diag  \( \diag \( \vec ^{-1} \( \bA \) \) \)  )$ for the $i$th row of the coefficient matrix $\bA$  corresponding to $\vec (\bPsi _{n-j})$, 
also $\be_n = \vech \( \vec^{-1} (  \bD_n) \)$ a.s.
Given the SV-It\^o model, volatility dynamics can be explained by the previous volatilities $\bPsi_k$'s and random fluctuation is modeled by the martingale difference term $\be_n$ which comes from the $\bZ_t$ process in \eqref{VAR}.
Furthermore, under some stationary conditions, the expectation of the daily integrated factor volatility is
	\begin{equation*}
	\E \left \{  \vech( \bPsi_n)  \right \} =  \( \bI _{r(r+1)/2} - \sum_{ j=1}^q \bbeta_j \)  ^{-1}  \bbeta_0 ,
	\end{equation*}
where the largest eigenvalue of $\sum_{j=1}^q \bbeta_j $ should be strictly less than 1. 

\begin{remark}
Recently, \citet{kim2019factor} introduced a factor GARCH-It\^o model to capture the market dynamics with high-frequency data and for a large number of assets. 
In specific, they imposed some GARCH-type dynamic structure on the eigenvalue process of the factor volatility matrix process, so the daily integrated eigenvalues are functions of historical squared factor returns. 
The proposed SV-It\^o model in this paper also imposes some dynamic AR structure on the factor volatility process so that the two models share similarities in their approaches. 
However, to explain the market dynamics at the low-frequency, the factor GARCH-It\^o model employs low-frequency factor return information while the SV-It\^o model uses the integrated factor volatilities over low-frequency periods. 
Our empirical study shows that incorporating integrated factor volatilities helps to capture the market dynamics promptly (see Section \ref{SEC-empirical}).
Moreover, the factor GARCH-It\^o model is restricted to the eigenvalue structure so that it cannot capture the cross-sectional dynamics.
On the other hand, the SV-It\^o model adopts a more general structure so that the dynamics for correlations among market factors can be modeled as well (see Section \ref{SEC-factor-vol}). 
\end{remark}


\section{Parameter estimation} \label{SEC-3}

In this section, we introduce estimation procedures for the factor loading and factor volatility matrices, as well as for the model parameteres, and establish their asymptotic properties.

\subsection{The model set-up and realized volatility matrix estimators}

Let $p$ be the total number of assets and $n$ be the total number of low-frequency observations. 
The high-frequency prices for the $i$th asset during the $k$th low-frequency period are observed at times $t_{i,k, \ell } \in (k-1, k]$, $i=1, \ldots, p$, $k=1, \ldots,n$ and $\ell=1, \ldots, m_{i,k}$. 
Denote $m_i$ the averaged number of high-frequency observations during each low-frequency period for the $i$th asset, that is, $m_i=\sum_{k=1}^n m_{i,k}/n$. 
Further let $m=\sum_{i=1}^p m_i/p$. 
Let $Y_{i, t_{i,k,\ell}}$ be the observed log price of the $i$th asset at time $t_{i,k,\ell}$. 
High-frequency data are normally non-synchronized so that $t_{i_1,k, \ell} \neq t_{i_2, k, \ell}$ for $i_1 \neq i_2$. 
Moreover, due to imperfections of the trading mechanisms \citep{ait2008high}, high-frequency data are often contaminated by market microstructure noises so that the observed log price $Y_{i,t_{i,k,\ell}}$ is a noisy version of the corresponding true log price $X_{i, t_{i,k,\ell}}$, that is, 
	\begin{equation}\label{observedPrice}
	Y _{i,   t_{i, k ,\ell}} = X_{i,  t_{i, k ,\ell}}  + \epsilon _{i, t_{i, k,\ell}}, \quad i=1,\ldots, p, k=1, \ldots, n, \ell =1, \ldots, m_{i,k},
	\end{equation}
where $\varepsilon_{i, t_{i,k,\ell}}$'s are stationary noises with mean zero and variance $\eta_i$.
We further assume that $\varepsilon_i$ and $X_i$ are independent with each other.

Given non-synchronized and noisy high-frequency data, researchers constructed nonparametric realized volatility matrix estimators that take advantage of sub-sampling and local-averaging techniques to remove the effect of market microstructure noises so that the integrated volatility matrix $\bGamma_k$ can be estimated consistently and efficiently.
Examples include multi-scale realized volatility matrix (MSRVM) \citep{zhang2011estimating}, pre-averaging realized volatility matrix (PRVM) \citep{christensen2010pre}, and kernel realized volatility matrix (KRVM) \citep{barndorff2011multivariate} estimators. 
See also \citet{ait2010high, bibinger2014estimating, fan2018robust, kim2018large, wang2010vast}. 
When the number of assets is $p$ finite, these estimators can achieve the optimal convergence rate of $m^{-1/4}$ in the presence of market microstructure noises for estimating $\bGamma_{k}$.


\subsection{Non-parametric factor volatility matrix estimation} \label{SEC-factor-vol}

When the number of assets $p$ is finite, we may view the factor process $\bff_t$ as the log price process $\bX_t$ and estimate the daily integrated factor volatility matrices $\bPsi_k$'s with the well-performing estimators such as the MSRVM, the PRVM, and the KRVM. 
The estimators of $\bPsi_k$'s and the AR structure described in \eqref{VAR-int} can be employed to estimate model parameter $\bbeta_j$'s directly in this case (see Sections \ref{SUBSEC-QMLE} and \ref{SUBSEC-LSE}). 
However, in practice, we often encounter a large number of assets and the factor process $\bff_t$ is latent. 
In this section, we first discuss how to estimate the latent factor volatility matrices $\bPsi_k$'s in the high-dimensional set-up.

Given the factor-based It\^o diffusion process in \eqref{model1}, the daily integrated volatility is
	\begin{equation*}
	\bGamma_{k} =\bL  \bPsi_k   \bL ^{\top} + \bGamma_{k} ^s \quad \text{ for } k=1, 2, \ldots, n.
	\end{equation*}
The idiosyncratic risk is related to the firm-specific risk so that the corresponding co-volatility matrices are sparse.  
In light of this, we impose the following sparse condition on the idiosyncratic volatility matrix $\bGamma_{k}^s = \(\Gamma_{k, ij} ^s \) _{i,j=1,\ldots, p}$: 
	\begin{equation} 
	\label{sparsity}
	\max_{k \in \NN} \max_{1\leq j \leq p} \sum_{i=1}^p |\Gamma_{k, ij}^s|  ^{\delta} |\Gamma_{k, ii}^s \Gamma_{k, jj}^s | ^{(1-\delta)/2} \leq M \pi(p) \quad \text{ a.s.},
	\end{equation}
where $\delta \in [0,1)$, $M$ is a positive bounded random variable and the sparsity level $\pi(p)$ diverges very slowly such as $\log p$.
On the other hand, the factor process $\bff_t$ often depends on a few common market factors such as industry sectors, inflation reports, Fed rate hikes, and oil prices. 
Thus, the number of market factors $r$ takes a much smaller value than the number of assets $p$, so we assume that the rank $r$ is finite. 
In this paper, we further assume that $r$ is known. 
For the latent factor model, we face the identifiability issue. 
To manage this issue, researchers often impose some structures on the factor loading matrix $\bL$ such as $\bL ^{\top} \bL=p \bI _r$ and also assume that the factor volatility matrices $\bPsi_k$'s are diagonal.
It follows that $\bL$ and $\bPsi_k$'s are corresponding to the eigenvectors and eigenvalues of the factor volatility matrices.   
Under these assumptions, \citet{kim2019factor} proposed the factor GARCH-It\^o model for the eigenvalues of the factor volatility matrices.
However, in this case, the correlation structure among assets is constant, which makes it difficult to investigate the cross-sectional market dynamics.

To account for the dynamics among assets, we consider the following structure that is more general. 
First, we assume that the idiosyncratic volatility matrix $\bGamma_k^s$ satisfies
	\begin{equation} \label{sparse_matrix}
	\bGamma_{k}^s  = \bGamma ^s \quad \text{a.s.} \quad \text{ for all } k=1,2,\ldots,n. 
	\end{equation}
Moreover, we assume that the factor volatility matrices $\bPsi_k$'s form a stationary process such that 
	\begin{equation*}
	\frac{1}{n} \sum_{k=1}^n   \bPsi_k ^2  - \( \frac{1}{n} \sum_{k=1}^n \bPsi_k \)^2 \to   \bPsi_{\infty}^2  \quad \text{ a.s.}
	\end{equation*} 
as $n$ goes to infinity, where the liming variable $ \bPsi_{\infty}^2 $ may be $\E \[ \left \{ \bPsi_1 -\E \( \bPsi_1\)\right \} ^2 \]$. 
For the factor loading matrix $\bL$, we assume that it satisfies $\bL ^{\top} \bL = p \bI_r$ and $\bL  \bV = \bL$, where $\bV$ denotes the eigenvectors of $\bPsi_{\infty}^2$.

Under these conditions, we propose the following procedure to estimate the factor loading and volatility matrices. 
First, we consider
	\begin{eqnarray*}
	\SS_n &=& \frac{1}{pn} \sum_{k=1}^n \(  \bGamma _k  - \bar{\bGamma}  \) ^2 =   \frac{1}{np} \sum_{k=1}^n \( \bL \bPsi_{k}  \bL^{\top}  - \bL \bar{\bPsi} \bL^{\top}\) ^2  \cr
	&=& \bL  \[  \frac{1}{n} \sum_{k=1}^n \( \bPsi_{k}      - \bar{\bPsi} \) ^2  \] \bL^{\top},
	\end{eqnarray*}
where $\bar{\bGamma}=\frac{1}{n} \sum_{k=1}^n \bGamma_{k}$ and $\bar{\bPsi} =\frac{1}{n} \sum_{k=1}^n  \bPsi_{k}$.
Note that $\SS_n$ is the rank $r$ matrix and is free of the idiosyncratic volatility matrix $\bGamma^s$. 
As $n \to \infty$, note that we have 
	\begin{equation*}
	p^{-2} \bL^{\top} \SS_n \bL \to \bPsi_{\infty}^2 \quad \text{ a.s.}
	\end{equation*}
Thus, the scaled factor loading matrix $p^{-1/2}\bL$ relates to the eigenvectors of $\SS_{\infty}= \bL \bPsi_{\infty}^2 \bL^{\top}$. 
We estimate $p^{-1/2}\bL$ by the first $r$ eigenvectors of $\SS_{n}$.
However, in practice, $\SS_n$ is not observable so that we use its estimator, 
	\begin{equation} \label{averaged-squaredvol}
	\hat{\SS}_{n,m}  =\frac{1}{np} \sum_{k=1}^n \( \hat{\bGamma}_k - \bar{\hat{\bGamma}} \) ^2, 
	\end{equation}
where $\bar{\hat{\bGamma}}  = \frac{1}{n} \sum_{k=1}^n \hat{\bGamma}_k$ and $\hat{\bGamma}_{k}$ is the MSRVM,  the PRVM, or  the KRVM estimator for daily integrated volatility matrix $\bGamma_{k}$. 
We then estimate the factor loading matrix $p^{-1/2} \bL$ by the first $r$ eigenvectors of $\hat{\SS}_{n,m}$ and denote the estimator as $p^{-1/2}\hat{\bL}$.
Finally, the factor volatility matrix estimators can be obtained by 
	\begin{equation*}
	\hat{\bPsi}_k = p^{-2}  \hat{\bL} ^{\top} \hat{\bGamma}_{k} \hat{\bL}, \quad k=1, \ldots, n.
	\end{equation*}

\begin{remark}
In the literature on approximate factor models, researchers often assume that the factor volatility matrices $\bPsi_k$'s are diagonal so that their entries are related to the eigenvalues. 
Our model structure includes the diagonal structure and when it is assumed, we do not need to require the stationary condition. 
\end{remark}

To investigate the asymptotic behaviors of the proposed non-parametric estimators, we need the following technical conditions.

\begin{assumption} \label{Assumption2}
	~
	\begin{enumerate}
		\item [(a)] For some given $b\geq 2$,  $ \sup_{k \in \NN} \max_{1\leq i,j \leq p} \E \(  |\hat{\Gamma}_{k, ij} -\Gamma_{k,ij} | ^{2b} \) \leq C m^{-b/2}$;
		
		\item [(b)] $ \E \( \|\frac{1}{n} \sum_{k=1}^n   \bPsi_k ^2  - \( \frac{1}{n} \sum_{k=1}^n \bPsi_k \)^2 -  \bPsi_{\infty}^2   \| _{F} ^b\) \leq C n^{-b/2}$;
		
		\item [(c)] For all $j=1,\ldots, r$, $\lambda_{j}( \SS_{\infty})- \lambda_{j+1} (\SS_{\infty}) \geq C p$ for some fixed constant $C$, where $\lambda_{j} (\bA)$ is the $j$th largest eigenvalue of the square matrix $\bA$. 
	\end{enumerate}
\end{assumption}

\begin{remark}
Assumption \ref{Assumption2} (a) is satisfied when the instantaneous volatility processes and noise have the finite $4b$th moment \citep{kim2016asymptotic, tao2013fast}. 
Assumption \ref{Assumption2} (b) is required to obtain the consistent estimator for $\bL$ that is uniquely defined.
Finally, Assumption \ref{Assumption2} (c) is the so-called pervasive condition which is often imposed when investigating the approximate factor models \citep{fan2013large}.
\end{remark}

We present the theorem that investigates the asymptotic behaviors of the proposed non-parametric estimators for the factor loading and factor integrated volatility matrix. 

\begin{thm} \label{Thm:non}
Under the models \eqref{model1} and \eqref{observedPrice}, when Assumption \ref{Assumption2}, \eqref{sparse_matrix}, and the sparsity condition \eqref{sparsity} are met, we have
	\begin{eqnarray}
	&& \E \( \| \hat{\SS}_{n,m} -\SS_{\infty} \|_F ^b \) \leq C p^b \( m^{-b/4} + n^{-b/2} \), \label{result1-Thm-non} \\
	&& \max_{1\leq i\leq r} \E \( \| p^{-1/2} \hat{\bL}_i   - p^{-1/2} sign (\hat{\bL}_i ^{\top} \bL_i)  \bL_i    \|_F ^b \) \leq C  \( m^{-b/4} + n^{-b/2} \), \label{result2-Thm-non} \\
	&& \sup_{k \in \NN} \E \( \| \hat{\bPsi}_{k} -\bPsi_{k} \|_F ^{2b/3} \) \leq C  \left \{ n^{-b/3} + m^{-b/6} + \( \pi(p) /p \) ^{2b/3} \right \},\label{result3-Thm-non}  
	\end{eqnarray}
where $\bL_i$ is the $i$th column of $\bL$.
\end{thm}

\begin{remark}
Theorem \ref{Thm:non} shows that the latent factor process can be estimated consistently with the convergence rate $n^{-1/2} + m^{-1/4}+\pi(p)/p$.
The term $n^{-1/2}$ is coming from identifying the latent factor loading matrix $\bL$ under the stationary condition. 
However, if we do impose the diagonal structure on the factor volatility matrix $\bPsi_k$, the term $n^{-1/2}$ is removed. 
The term $m^{-1/4}$ is coming from estimating the daily integrated volatility matrix $\bGamma_k$, which is known as the optimal convergence rate in the presence of the market microstructure noises. 
Finally, the term $\pi(p)/p$ is the cost to identify the latent factor volatility matrix $\bPsi_k$ from the integrated volatility matrix $\bGamma_k$. 
These results imply Assumption \ref{Assumption1} (d) and helps to establish the convergence rate in Theorem \ref{Thm:theta}. 
\end{remark}


\subsection{Quasi-maximum likelihood estimation}
\label{SUBSEC-QMLE}

In this section, we propose a quasi-maximum likelihood estimation procedure for the true parameter $\btheta_0 =(\bbeta_{0,0}^{\top}, \vec(\bbeta_{0,1})^{\top},\ldots, \vec (\bbeta_{0,q})^{\top})^{\top} \in \RR ^d$, 
where $\bbeta_j$'s are defined in \eqref{VAR-int}  and $d=r (r +1)\{  2+q r (r+1) \}/4$.

We first develop the estimation procedure by pretending that $\bPsi_k$'s are known and consider the following quasi-likelihood function: 
	\begin{equation*}
	L_{n} (\btheta) =- \frac{1}{n} \sum_{k=q+1}^n  \[ \log \( \det (\bH_k(\btheta))\) +\Tr\(\bPsi_k \bH_k^{-1}(\btheta) \) \], 
	\end{equation*}
where $\vech \( \bH_k (\btheta) \) = \bbeta_0  + \sum_{j=1}^q \bbeta_j \vech(\bPsi_{k-j})$ and $\bH_k (\btheta) $ is symmetric. 
The difference between $\bPsi_k$ and $\bH_k (\btheta_0)$ under the SV-It\^o model is the martingale difference whose vectorization is $\be_n$ defined in \eqref{VAR-int}.
Then, under some technical conditions, we can show that the maximizer of $L_{n} (\btheta)$ converges to the true parameter $\btheta_0$ with the convergence rate of $n^{-1/2}$.
However, the daily integrated factor volatility matrix $\bPsi_k$'s are unobservable so that we need to follow the procedure developed in Section \ref{SEC-factor-vol} to obtain their estimators $\hat{\bPsi}_k$'s.
Given the estimators $\hat{\bPsi}_k$'s, we let 
	\begin{equation}
	\vech \( \hat{\bH}_k (\btheta) \) = \bbeta_0  + \sum_{j=1}^q \bbeta_j \vech(\hat{\bPsi}_{k-j}), 
	\label{Vol-est}
	\end{equation}
where $\hat{\bH}_k (\btheta)$ is symmetric, and define the following quasi-likelihood function for parameter estimation: 
	\begin{equation}
	\hat{L}_{n,m} (\btheta) = - \frac{1}{n} \sum_{k=q+1}^n  \[ \log \( \det (\hat{\bH}_k(\btheta))\) +\Tr\(\hat{\bPsi}_k \hat{\bH}_k^{-1}(\btheta) \) \].
	\label{quasi-likelihood function}
	\end{equation}
The true model parameters $\btheta_0$ can be obtained by maximizing the quasi-likelihood function in \eqref{quasi-likelihood function}, that is, 
	\begin{equation*}
	\hat{\btheta}  = \arg \max_{\btheta \in \bTheta} \hat{L}_{n,m} (\btheta),
	\end{equation*} 
where $\bTheta$ is the parametric space of $\btheta$.

To investigate the asymptotic behaviors of the proposed QMLE, we need the following technical conditions.

\begin{assumption} \label{Assumption1}
	~
	\begin{enumerate}
		\item [(a)] $\bTheta$ is compact;  $\btheta_0$ is an interior point of $\bTheta$; 
		
		\item [(b)] $\min_{k \in \NN} \inf_{\btheta \in \bTheta} \lambda_{\min} (\bH_k (\btheta)) > c_\lambda $ and $\min_{k \in \NN} \inf_{\btheta \in \bTheta} \lambda_{\min} (\hat{\bH}_k (\btheta)) > c_\lambda$ a.s. for some fixed positive constant $c_{\lambda}$, $\lambda_{\min} (\bA)$ is the smallest eigenvalue of a square matrix $\bA$;
	
		\item[(c)] There exists some fixed constants $C_1$ and $C_2$ such that $C_1 m_i \leq m_{i,k} \leq C_2 m_i$ for all $i$ and $k$, and $C_1 m \leq m_i \leq C_2 m$ for all $i$. 
		
		\item [(d)] There is some fixed sequence $\tau_m$ such that 
		\begin{equation*}
		\sup_{k \in \NN}  \E \(   \| \hat{\bPsi}_k -\bPsi_k  \|_F^4 \) \leq \tau_m ^{4} =o(1). 
		\end{equation*} 
	\end{enumerate}
\end{assumption}

\begin{remark}
Assumption \ref{Assumption1} (a) is required to define the parameter uniquely.
Assumption \ref{Assumption1} (b) is often obtained by imposing some positive definitiveness requirement on the intercept part $\bbeta_0$.
Assumption \ref{Assumption1} (d) is required to investigate the asymptotic behavior of QMLE $\hat{\btheta}$ since the latent factor volatility matrices $\bPsi_k$'s are not observable, and we need to estimate them. 
When the number of assets $p$ is fixed, a simple way to establish the rate $\tau_m$ is to treat the factor process $\bff_t$ as the log price process $\bX_t$, then the integrated volatility matrices $\bPsi_k$'s can be estimated by the MSRVM, the PRVM, or the KRVM estimator.
These well-performing estimators can achieve the optimal convergence rate of $m^{-1/4}$ in estimating $\bPsi_k$'s when observed stock prices are contaminated by market microstructure noises. 
$\tau_m$ follows to be $m^{-1/4}$ in this case. 
On the other hand, in the high-dimensional set-up, we need to identify the latent factor volatility matrix and are required to impose some structures on the factor loading matrix so that $\tau_m$ can be appropriately established. 
In this case, the estimation procedure presented in Section \ref{SEC-factor-vol} has $\tau_m=n^{-1/2} + m^{-1/4}+\pi(p)/p$ (see Theorem \ref{Thm:non}). 
\end{remark}

The following theorem provides the convergence rate of the QMLE $\hat{\btheta}$.
\begin{thm} \label{Thm:theta}
Under Assumption \ref{Assumption1},  we have 
	\begin{equation} \label{result1-Thm-theta}
	\left \| \hat{\btheta}   -\btheta_0 \right \|_{\max} = O_p (\tau_m + n^{-1/2}).
	\end{equation}
\end{thm}

\begin{remark}
Theorem \ref{Thm:theta} shows that the convergence rate of the QMLE $\hat{\btheta}$ is $\tau_m + n^{-1/2}$.
The first term $\tau_m$ is the cost to estimate the integrated factor volatility matrices $\bPsi_k$'s and takes the value $m^{-1/4}+ n^{-1/2} + p/\pi(p)$ (see Section \ref{SEC-factor-vol}).
The second term $n^{-1/2}$ is the usual convergence rate for parametric estimation based on the low-frequency structure. 
The asymptotic result obtained in Theorem \ref{Thm:theta} holds as long as the daily integrated factor volatility matrices $\bPsi_k$'s satisfy \eqref{VAR-int}.
That is, the asymptotic result does not depend on the specific instantaneous volatility process described in \eqref{VAR}. 
For example, we can develop an instantaneous volatility process that can capture intraday volatility dynamics such as the U-shape pattern \citep{admati1988theory, andersen1997intraday, andersen2018time, hong2000trading} in the following: 
	\begin{eqnarray*}   
	\bSigma_t  &=& \bSigma _{[t]} + (t-[t])^2 \balpha_0 ^{\prime}\(  \bI_r  + \sum_{j=1}^{q-1} \balpha_{j+1} \bPsi _{[t]-j+1} \balpha_{j+1} ^{  \top} \) \balpha_0 ^{\prime \top}  \cr
	&&	- (t-[t]) \( \balpha_0 \balpha_0 ^{\top} + \bSigma _{[t]}  + \sum_{j=1}^{q-1} \balpha_{j+1} \bPsi _{[t]-j+1} \balpha_{j+1} ^{\top} \)  \cr
	&&+ \balpha_{1} \(\int_{[t]} ^{t} \bSigma_s ds \) \balpha_1 ^{\top} + ([t] +1 -t) \bZ _t \bZ_t^{\top}.
	\end{eqnarray*}
We then still obtain the AR structure as described in \eqref{VAR-int}. 
If consistent estimators for the instantaneous volatility $\bSigma_t$'s were available, we could also study the intraday dynamics.
However, this is not the focus of this paper so that we leave it for future research. 
\end{remark}


\subsection{Least squares estimation}
\label{SUBSEC-LSE}

When considering the vector autoregression form \eqref{VAR-int} directly, one of the natural ways to estimate the parameters is the well-known least squares estimation method. 
More specifically, we define the square loss function as: 
	\begin{equation*}
	\hat{L}_{n,m}^{ls} (\btheta) =   \frac{1}{n} \sum_{k=q+1}^n  \| \vech (\hat{\bPsi}_k) - \bbeta_0  - \sum_{j=1}^q \bbeta_j \vech(\hat{\bPsi}_{k-j})  \|_2^2.
	\end{equation*}
The true model parameters $\btheta_0$ can be obtained by minimizing the above square loss function, that is, 
	\begin{equation*}
	\hat{\btheta}^{ls}  = \arg \min_{\btheta \in \bTheta} \hat{L}_{n,m} ^{ls}(\btheta).
	\end{equation*} 
We call the estimator the LSE.
Its asymptotic behavior can be showed similar to the proofs of Theorem \ref{Thm:theta}, and the convergence rate is 
	\begin{equation*}
	\left \| \hat{\btheta}  ^{ls} -\btheta_0 \right \|_{\max} = O_p (\tau_m + n^{-1/2}).
	\end{equation*}
It is easy to implement the LSE method as it has the closed form which is the well-known ordinary least squares estimator. 
However, the vector autoregression equation \eqref{VAR-int} results in heterogeneous noise described by the term $\be_n$, 
which may cause inefficiency in estimating model parameters using the LSE method. 
On the other hand, the QMLE method adjusts the heterogeneous variance term using the inverse of the conditional variance matrix function $\hat{\bH}^{-1} (\btheta)$. 
However, the implementation of the QMLE is demanding and its performance may depend on the choice of the initial value of the optimization. 
To overcome these issues, we suggest adopting the LSE as the initial value for the QMLE procedure. 


\section{Large volatility matrix prediction}\label{SEC-4}

In finance applications such as portfolio allocation, we are often required to predict future large volatility matrix. 
In this section, we demonstrate how to harness the proposed SV-It\^o model for constructing a predictor. 
It is known that the best predictor for future large volatility matrix is the conditional expected value of the integrated volatility matrix given current information. 
Under the SV-It\^o model and the condition \eqref{sparse_matrix}, we have 
	\begin{equation*}
	\E \( \bGamma_{n+1} \middle | \FF_{n} \) = \bL \bH_{n+1} (\btheta_0)  \bL^{\top} + \bGamma ^s \quad \text{a.s.}
	\end{equation*}
That is, the conditional expectation consists of the conditional expected factor volatility and idiosyncratic volatility matrices.

We first discuss the estimation of the idiosyncratic volatility matrix. 
Note that  	
	\begin{eqnarray*} 
	\bar{\bGamma} &=& \bL \(\frac{1}{n}  \sum_{k=1}^{n} \bPsi _k \) \bL^{\top} + \bGamma ^s \cr
	&=&  \bPhi_n+ \bGamma^s,
	\end{eqnarray*}
where the idiosyncratic volatility matrix $\bGamma^s$ satisfies the sparsity condition \eqref{sparsity}, and the averaged factor volatility matrix $\bPhi_n=\bL \(\frac{1}{n}  \sum_{k=1}^{n} \bPsi _k \) \bL^{\top}$ has the finite rank $r$. 
Therefore, the averaged integrated volatility matrix $\bar{\bGamma}$ also retains a low-rank plus sparse structure. 
Given this structure, we can apply the POET procedure introduced by \citet{fan2013large} to estimate the idiosyncratic volatility matrix $\bGamma^s$.
More specifically, the input of idiosyncratic volatility estimator is 
	\begin{equation*}
	\tilde{\bGamma}^s= (\tilde{\Gamma}_{ij}^s)_{i,j=1,\ldots,p}  =\bar{\hat{\bGamma}}- \sum_{j=1}^r \hat{\blambda}_j \hat{\bq}_j  \hat{\bq}_j ^{\top}, 
	\end{equation*}
where  $\hat{\lambda}_j$ is the $j$th largest eigenvalue of $\bar{\hat{\bGamma}}$ and $\hat{\bq}_j$ is the corresponding eigenvector.
We then apply the adaptive threshold scheme to the input of idiosyncratic volatility matrix estimator as follows:
	\begin{eqnarray*}
	\hat{\Gamma}_{ij}^s= 
		\begin{cases}
		s_{ij} (\tilde{\Gamma}_{ij} ^s)  \1 ( | \tilde{\Gamma}_{ij} ^s |\geq \varpi_{ij} ) & \text{ if } i \neq j  \\ 
		\tilde{\Gamma}_{ij}^s \vee 0 & \text{ if } i=j, 
		\end{cases}
		\quad \text{and} \quad  \hat{\bGamma}^s = (\hat{\Gamma}_{ij}^s) _{i,j=1, \ldots, p}, 
	\end{eqnarray*} 
where the thresholding function $s_{ij} (\tilde{\Gamma}_{ij}^s)$ satisfies $|s_{ij} (\tilde{\Gamma}_{ij}^s) - \tilde{\Gamma}_{ij} ^s | \leq \varpi_{ij}$,
and we use the adaptive threshold level $\varpi_{ij}= \varpi _m \sqrt{(\tilde{\Gamma}_{ii} ^s\vee 0 )(\tilde{\Gamma}_{jj}^s \vee 0 )} $ which is the same as applying the threshold $\varpi_m$ to the correlation.

On the other hand, with the QMLE $\hat{\btheta}$, we estimate the factor volatility matrix by 
	\begin{equation*}
	\hat{\bL} \hat{\bH}_{n+1} (\hat{\btheta})  \hat{\bL}^{\top},
	\end{equation*}
where $\hat{\bL}$ is estimated by the first $r$ eigenvectors of $\hat{\SS}_{n,m}$ defined in \eqref{averaged-squaredvol} and the AR structure for daily integrated volatility matrices $\hat{\bH}_k (\btheta)$ is defined in \eqref{Vol-est}.
Combining the idiosyncratic volatility matrix estimator $\hat{\bGamma}^s$, we estimate the future large volatility matrix by
	\begin{equation*}
	\tilde{\bGamma}_{n+1} = \hat{\bL} \hat{\bH}_{n+1} (\hat{\btheta})  \hat{\bL}^{\top} + \hat{\bGamma}^s.
	\end{equation*}
We name the proposed estimator the SV-It\^o POET (SV-POET) estimator.

To investigate the asymptotic behaviors of the proposed estimator, we need the following assumptions.

\begin{assumption} \label{Assumption3}
	~
	\begin{enumerate}
		\item [(a)]  For some fixed positive constant $C_1$, we have
		\begin{equation*}
		\frac{p}{r} \max_{1\leq  i \leq p} \sum_{j=1}^r q_{ij} ^2 \leq C_1 \text{ a.s.},
		\end{equation*}
		where $\bq_j = (q_{1j}, \ldots, q_{pj})^{\top}$ is the eigenvector of the averaged factor volatility matrix $\bPhi_n$ corresponding to the  $j$th largest eigenvalue;
		
		\item[(b)] We have $D_{\lambda} \geq C_2 p$ and $\lambda_1 / D_\lambda \leq C_3 $ a.s., where $D_\lambda= \min \{\bar{ \lambda}_i -\bar{\lambda}_{i+1}: i=1,\ldots,r \}$, $\bar{\lambda}_i$ is the $i$th largest eigenvalue of $\bPhi_n$, $\lambda_1$ is the largest eigenvalue of $\bar{\bGamma}$, 
		and the smallest eigenvalue of $\bGamma^s$ stays away from zero; 
		
		\item [(c)] $\pi(p) /p^{1/2} + \sqrt{ \log p / (n m^{1/2}+m) }= o(1)$.
	\end{enumerate}
\end{assumption}

\begin{remark}
Assumption \ref{Assumption3} (a) and (b) are called the incoherence condition and pervasive condition, respectively, which are often imposed in analyzing low-rank matrix and approximate factor models  \citep{ait2017using, candes2011robust, fan2013large}.
\end{remark}

The following theorem investigates the asymptotic behaviors of the SV-POET. 

\begin{thm} \label{Thm-future}
Under the models \eqref{model1}, \eqref{VAR}, and \eqref{observedPrice}, the following concentration inequality, 
	\begin{equation}\label{condition-Thm-future}
	\Pr \left \{\max_{1 \leq i,j \leq p} | \bar{\hat{\Gamma}}_{ij} -\bar{\Gamma}_{ij} | \geq C \sqrt{\frac{\log p}{m^{1/2} n +m}}  \right \} \leq p^{-1},
	\end{equation}
Assumptions \ref{Assumption2}--\ref{Assumption3}, \eqref{sparse_matrix}, and the sparsity condition \eqref{sparsity} are met. 
Take $\varpi_m = C_{\varpi} ( \pi(p) /p + \sqrt{\log p / ( n m^{1/2 } +m ) } )$ for some large fixed constant $C_{\varpi}$, then we have 
	\begin{eqnarray}
	&& \|\hat{\bGamma} ^s - \bGamma ^s \| _2 = O_p ( \pi (p) \varpi_m ^{1-\delta} ), \label{result1-Thm-future}\\
	&& \|\hat{\bGamma} ^s - \bGamma ^s \| _{\max} = O_p (  \varpi_m ), \label{result2-Thm-future}\\
	&& \|\tilde{\bGamma} _{n+1}  - \bGamma^* \| _{\bGamma^*} = O_p  \Big (   \tau_m +   n^{-1/2} +  m^{-1/4} + p^{1/2}  (\tau_m ^2+   n^{-1} +  m^{-1/2}  ) \cr
	&&\qquad \qquad \qquad \qquad \qquad   + \pi(p) \varpi_m ^{1-\delta}\Big )  , \label{result3-Thm-future} 
	\end{eqnarray} 
	where $\bGamma^* = \E \(\bGamma_{n+1}  \middle | \FF_n \)$, and the relative Frobenius norm is $\| \bA -  \bGamma^* \| _{\bGamma^*}  = p^{-1/2} \| \bGamma^{*-1/2} (\bA -  \bGamma^*  ) \bGamma^{*-1/2} \|_F$.
\end{thm}

\begin{remark}
The concentration inequality condition \eqref{condition-Thm-future} has the convergence rate $n^{-1/2} m^{-1/4}+ m^{-1/2}$ which is faster than the usual convergence rate $m^{-1/4}$.
The reason is as follows.
Usually, we investigate the asymptotic behavior with finite sample period, that is, $n$ is not allowed to go to infinity, and so the convergence rate merely depends on the high-frequency sample size $m$. 
However, in our setting, we allow the low-frequency sample size $n$ to go to infinity as well. 
Then the low-frequency summation employs some martingale structure, and we therefore can enjoy the faster convergence rate $n^{-1/2} m^{-1/4}$.
The additional term $m^{-1/2}$ is coming from some non-martingale terms such as the drift term.
To obtain the sub-Gaussian concentration inequality, we need some sub-Gaussian condition on the observed log stock prices $\bY_t$ such as the bounded instantaneous volatility condition  \citep{tao2013optimal}. 
Recently, \citet{fan2018robust} proposed the robust pre-averaged volatility estimation scheme and the sub-Gaussian concentration inequality can be obtained even when the observed log stock prices are heavy-tailed.
Thus, this condition is not restrictive.
\end{remark}

\begin{remark}
Theorem \ref{Thm-future} shows that the estimator for future large volatility matrix, $\tilde{\bGamma}_{n+1}$, has the convergence rate of $\tau_m +   n^{-1/2} +  m^{-1/4} + p^{1/2}  (\tau_m ^2+   n^{-1} +  m^{-1/2}  ) + \pi(p) \varpi_m ^{1-\delta}$.
Note that $\tau_m$ depends on the non-parametric estimators in Section \ref{SEC-factor-vol} and $\tau_m$ may be $n^{-1/2} + m^{-1/4} + \pi(p) /p$.  
In this case, the convergence rate will be $n^{-1/2} +  m^{-1/4} +\pi(p)/p + p^{1/2}  (  n^{-1} + m^{-1/2}   ) + \pi(p) \varpi_m ^{1-\delta}$ and the SV-POET estimator $\tilde{\bGamma}_{n+1}$ is consistent as long as $p= o (n^2)$ and $p= o(m)$. 
\end{remark}



\section{Numerical analysis} \label{SEC-5}

\subsection{A simulation study}
\label{SEC-simulation}

In this section, a simulation study was conducted to check finite sample performance of the proposed parameter estimators $\hbtheta$ and $\hbtheta^{ls}$, as well as to investigate the prediction performance of the proposed SV-POET estimator $\tilde{\bGamma}$, which was also compared with the performance of the estimator for future large volatility matrix proposed in \citet{kim2019factor}. 
Let $p$ be the total number of assets studied, $n$ be the total number of low-frequency observations, and $m$ be the total number of high-frequency observations during each low-frequency period. 
Log prices $\bX_t=(X_{1,t}, \ldots, X_{p,t})^{\top}$ at discrete time points $t_{i,j}=i-1+j/m$,   $i=1, \ldots, n$ and $j=1, \ldots, m$, were generated according to \eqref{model1} with $\bmu_t=0$ by the Euler scheme. 
Standard Brownian motions such as $\bB_t$ and $\bW_t$ were simulated by the normalized partial sums of independent standard normal random variables.
We considered a scenario where $r=3$ that suggests three market factors exert an impact on all trading stocks. 
For the instantaneous factor volatility process in \eqref{VAR}, we considered a diffusion process with $q=1$ so that when the continuous process is restricted to integer times, it retains an AR(1) structure.
For the parameters in \eqref{VAR}, we took the following set of values within this simulation study: 
$\balpha_0=\Diag(0.5,0.4,0.3)$, $\vec(\balpha_1)=(0.2,0,0,0.5,0.5,-0.2,0.8,-0.5,0.3)^\top$, $\nu=\Diag(0.5,0.5,0.5)$. 
This set of model parameters results in the following target parameters for estimation:
	\begin{equation*}
	\bbeta_{0,0}^{\top}=\left( 0.367,0,0.005,0.252,-0.024,0.143 \right),
	\end{equation*}
	\begin{equation*}
	\bbeta_{0,1}= \left(
	\begin{array}{cccccc}
	0.021 & 0.105 & 0.164 & 0.138 & 0.418 & 0.328 \\ 
	0 & 0.055 & -0.056 & 0.150 & 0.063 & -0.219 \\ 
	0 & -0.022 & 0.033 & -0.062 & 0.001 & 0.129 \\ 
	0 & 0 & 0 & 0.175 & -0.365 & 0.191 \\ 
	0 & 0 & 0 & -0.073 & 0.179 & -0.106 \\ 
	0 & 0 & 0 & 0.031 & -0.085 & 0.060 \\ 
	\end{array}
	\right).
	\end{equation*}
Initial values for the instantaneous factor volatility process were $\E \[ \bPsi_k \]$. 
The factor loading matrix $\bL$ is a $p$-by-$r$ matrix, where the first column takes values $\sqrt{2} \cos \left( 2i \pi/p \right)$, $i=1, \ldots, p$, the second column takes values $\sqrt{2} \sin \left( 2i \pi/p \right)$, $i=1, \ldots, p$, and the third column entries share the same value 1 so that the factor loading matrix retains the structure such that $\bL^{\top} \bL=p \bI_r$.
On the other hand, to generate the idiosyncratic diffusion process that has a sparse structure in its daily integrated co-volatility $\bGamma^s$, we took $\bGamma^s=\left(\Gamma^s_{ij} \right)_{1 \leq i,j \leq p}$, where
	\begin{equation*}
	\Gamma^s_{ij} =0.1 \cdot 0.5^{ |i-j| } \sqrt{\Gamma_{ii} \Gamma_{jj}}
	\end{equation*}
for the off-diagonal elements and $\Gamma^s_{ii}=0.1, i=1, \ldots, p$, for the diagonal elements. 
For the high-frequency data $Y_{i,t}$ observed between integer times, we added market microstructure noises to the simulated log price $X_{i,t}$ where the noises were modeled by independent normal random variables with mean 0 and standard deviation $0.005$.
Given the simulated log prices $Y_{i,t}$, we employed the PRVM \citep{christensen2010pre} estimator to obtain daily integrated volatility matrix estimator $\hbGamma_k$, $k=1,\ldots,n$. 
The sample variance of PRVM estimators, $\hat{\SS}_{n,m}$, was then computed and its first $r=3$ eigenvectors were adopted to estimate factor loading matrix $\bL$. 
Parameter matrices $\bbeta_{0}$ and $\bbeta_{1}$ were estimated by either maximizing the proposed likelihood function $\hat{L}_{n,m} (\btheta)$ or minimizing the proposed loss function $\hat{L}^{ls}_{n,m} (\btheta)$. 
Parameter estimates from the LSE method were used to initialize the optimization algorithm for the QMLE method. 
We took $n=125,250,500$ and $m=390,780,2340$ with $p=200$. 
For each combination of $n$ and $m$, we repeated the simulation for 500 times.

Tables \ref{table: beta0} and \ref{table: beta1} summarize the mean spectral norms, Frobenius norms, and max norms of $\hbbeta_0 - \bbeta_{0}$ and $\hbbeta_1 - \bbeta_{1}$ given both the QMLE and LSE methods. 
The results show that as the number of low-frequency or high-frequency observations increases, the estimation performance becomes better, which support the theoretical results derived in Section \ref{SEC-3}. 
Moreover, the QMLE method provides more accurate estimation results than the LSE method. 
The underlying reason may be that the QMLE method is capable of adjusting the heterogeneous volatility. 

\begin{table}
	\centering
	\begin{tabular}{cccccccccc}
		\hline 
		\hline 	
		&& \multicolumn{2}{c}{\bf{Spectral Norm}} && \multicolumn{2}{c}{\bf{Frobenius Norm}} && \multicolumn{2}{c}{\bf{Max Norm}} \\
		\cline{3-4} \cline{6-7}\cline{9-10}  
		$n$&$m$& QMLE & LSE && QMLE & LSE && QMLE & LSE \\
		\hline 
		\multirow{3}{*}{125} & 390 & 0.156 & 0.158 && 0.156 & 0.158 && 0.119 & 0.120 \\
		& 780  & 0.147 & 0.149 && 0.147 & 0.149 && 0.117 & 0.118 \\
		& 2340 & 0.140 & 0.142 && 0.139 & 0.140 && 0.115 & 0.116 \\
		&&&&&&&&&\\
		\multirow{3}{*}{250} & 390 & 0.139 & 0.141 && 0.139 & 0.141 && 0.110 & 0.112 \\
		& 780  & 0.132 & 0.135 && 0.132 & 0.135 && 0.107 & 0.110 \\
		& 2340 & 0.129 & 0.130 && 0.129 & 0.130 && 0.105 & 0.109 \\
		&&&&&&&&&\\
		\multirow{3}{*}{500} & 390 & 0.132 & 0.133 && 0.131 & 0.132 && 0.106 & 0.107 \\
		& 780 & 0.126 & 0.127 && 0.126 & 0.127 && 0.105 & 0.105 \\
		& 2340 & 0.120 & 0.121 && 0.120 & 0.121 && 0.103 & 0.104 \\
		\hline 		
		\hline 
	\end{tabular}
	\caption{The mean spectral norms, Frobenius norms, and max norms of $\hbbeta_0 - \bbeta_{0}$ for $p=200$, $n=125,250,500$ and $m=390, 780, 2340$.
	\label{table: beta0}}
\end{table}

\begin{table}
	\centering
	\begin{tabular}{cccccccccc}
		\hline 
		\hline 	
		&& \multicolumn{2}{c}{\bf{Spectral Norm}} && \multicolumn{2}{c}{\bf{Frobenius Norm}} && \multicolumn{2}{c}{\bf{Max Norm}} \\
		\cline{3-4} \cline{6-7}\cline{9-10}  
		$n$&$m$& QMLE & LSE && QMLE & LSE && QMLE & LSE \\
		\hline 
		\multirow{3}{*}{125} & 390 & 0.810 & 0.830 &  & 1.042 & 1.068 &  & 0.534 & 0.544 \\ 
		& 780  & 0.797 & 0.804 &  & 1.012 & 1.025 &  & 0.533 & 0.542 \\ 
		& 2340 & 0.791 & 0.800 &  & 1.009 & 1.018 &  & 0.532 & 0.537 \\ 
		&&&&&&&&&\\
		\multirow{3}{*}{250} & 390 & 0.733 & 0.743 &  & 0.921 & 0.938 &  & 0.476 & 0.481 \\ 
		& 780  & 0.702 & 0.710 &  & 0.880 & 0.889 &  & 0.471 & 0.475 \\ 
		& 2340 & 0.689 & 0.693 &  & 0.872 & 0.880 &  & 0.467 & 0.469 \\ 
		&&&&&&&&&\\
		\multirow{3}{*}{500} & 390 & 0.700 & 0.715 &  & 0.869 & 0.884 &  & 0.477 & 0.486 \\ 
		& 780 & 0.664 & 0.669 &  & 0.823 & 0.825 &  & 0.454 & 0.458 \\ 
		& 2340 & 0.644 & 0.645 &  & 0.791 & 0.797 &  & 0.442 & 0.447 \\ 
		\hline 		
		\hline 
	\end{tabular}
	\caption{The mean spectral norms, Frobenius norms, and max norms of $\hbbeta_1 - \bbeta_{1}$ for $p=200$, $n=125,250,500$ and $m=390, 780, 2340$.
	\label{table: beta1}}
\end{table}

The major motivation of our model proposal is to predict future large volatility matrix by taking advantage of the imposed AR model structure at the low-frequency. 
So we examined the finite sample performance of the proposed estimator $\tilde{\bGamma}_{n+1}$ for the conditional integrated volatility matrix $\E \( \bGamma_{n+1} \middle | \FF_{n} \)$ based on the procedure described in Section \ref{SEC-4}.
When estimating the idiosyncratic volatility matrix $\bGamma^s$, we applied the threshold $\sqrt{2 \log p/ (n m^{1/2}+ m) }$ on its input $\tilde{\bGamma}^s$. 
For each simulation, we computed the matrix estimation errors in spectral, max, and relative Frobenius norms respectively: 
	\begin{equation*}
	\begin{split}
	& \| \tilde{\bGamma}_{n+1} - \E \( \bGamma_{n+1} \middle | \FF_{n} \) \|_2 / \| \E \( \bGamma_{n+1} \middle | \FF_{n} \) \|_2, \\
	& \| \tilde{\bGamma}_{n+1} - \E \( \bGamma_{n+1} \middle | \FF_{n} \) \|_{\max} / \| \E \( \bGamma_{n+1} \middle | \FF_{n} \) \|_{\max}, \\
	& \| \tilde{\bGamma}_{n+1} - \E \( \bGamma_{n+1} \middle | \FF_{n} \) \|_{\E \( \bGamma_{n+1} \middle | \FF_{n} \)}.
	\end{split}
	\end{equation*}
For comparison purpose, we as well examined the prediction performance of the factor and aggregated factor GARCH-It\^o model proposed by \citet{kim2019factor}. 
In specific, \citet{kim2019factor} modeled the eigenvalues of factor volatility matrices by some GARCH-type structure. 
They also proposed to estimate the factor loading matrix $\bL$ in some aggregated form and named the corresponding model as aggregated factor GARCH-It\^o model.
For both models, we used $r=3$ and applied threshold $\sqrt{2 \log p/ m^{1/2}}$ for the idiosyncratic volatility matrix estimation. 
On the other hand, $\E \( \bGamma_{n+1} \middle | \FF_{n} \)$ has the structure of low-rank plus sparse, thus, we considered the POET procedure introduced by \cite{fan2013large} to account for such structure. 
In specific, we chose threshold $\sqrt{2 \log p/ m^{1/2}}$ and used the POET estimator from the previous period $\hat{\bGamma}^{POET}_n$ to estimate $\E \( \bGamma_{n+1} \middle | \FF_{n} \)$ since when the parametric models are not considered, we often assume martingale structure instead. 
For the benchmark, we also considered the PRVM estimator $\hat{\bGamma}_n$ from the previous period.

\begin{figure}
\centering
\includegraphics[width=\textwidth]{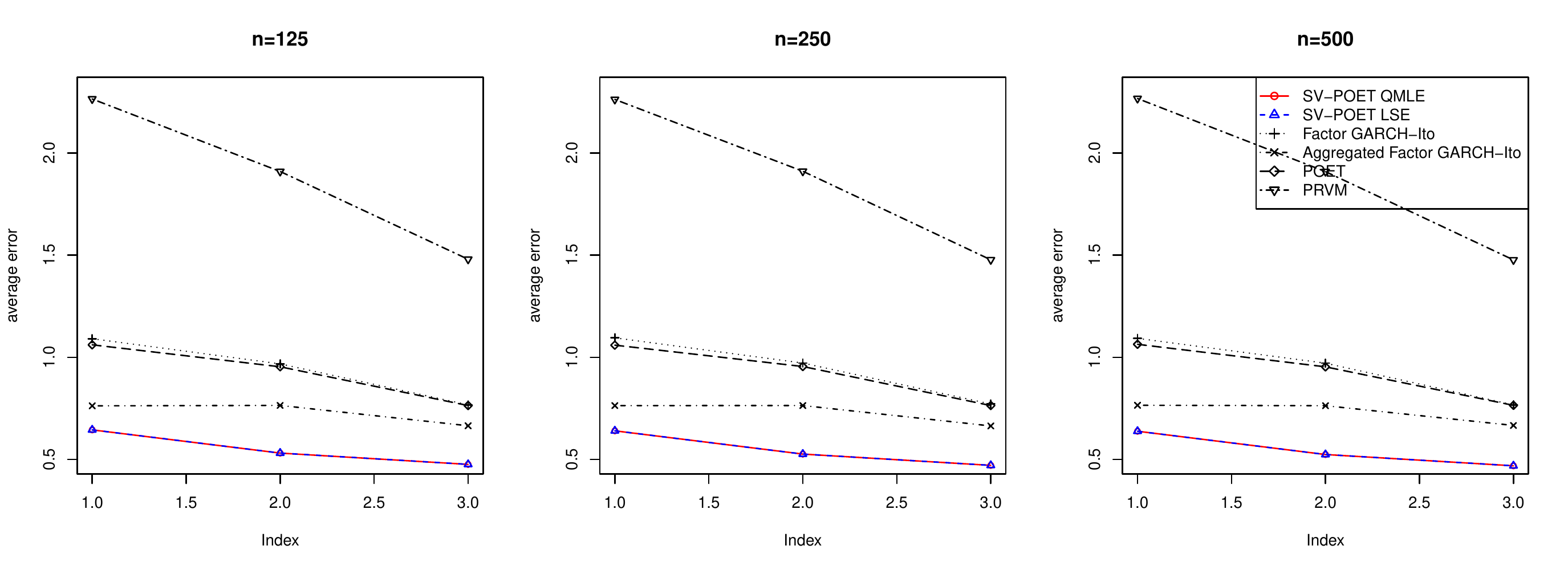}
\vspace*{-0.5cm} 
\caption{Mean matrix estimation errors in the relative Frobenius norms for the conditional daily integrated volatility matrix against $m$ under different $n$ values. \label{figure: forecast}}
\end{figure}

Table \ref{table: matrix estimation} summarizes the mean matrix estimation errors in the spectral, max, and relative Frobenius norms while Figure \ref{figure: forecast} plots the mean estimation errors in the relative Frobenius norms against the number, $m$, of high-frequency observations. 
The proposed SV-POET estimator outperforms the factor and aggregated factor GARCH-It\^o, the POET, and the PRVM methods. 
The QMLE method provides a bit more accurate prediction results than the LSE method. 
As the number of low-frequency or high-frequency observations increases, the mean estimation errors decrease for the SV-POET method, which supports the theoretical results in Section \ref{SEC-4}. 
Moreover, the prediction performance of the POET and PRVM only consistently improve given an increasing number of high-frequency observations. 
This may be because that these estimators are obtained using only the previous period high-frequency observations.

\begin{table}
	\small{
		\centering
		\begin{tabular}{ccccccccc}
			\hline 
			\hline 
			&&& \multicolumn{6}{c}{\textbf{Mean matrix estimation errors} $\mathbf{\times 100}$ } \\
			\cline{4-9}
			\multirow{3}{*}{\shortstack[c] {\textbf{Matrix} \\ \textbf{norms}}} & \multirow{3}{*}{$n$} & \multirow{3}{*}{$m$} & \multicolumn{2}{c}{\multirow{2}{*}{\textbf{SV-POET}}} & \multirow{3}{*}{\shortstack[c] {\textbf{Factor} \\ \textbf{GARCH-It\^o}}} & \multirow{3}{*}{\shortstack[c] {\textbf{Aggregated} \\ \textbf{factor} \\ \textbf{GARCH-It\^o}}} & \multirow{3}{*}{\textbf{POET}} &  \multirow{3}{*}{\textbf{PRVM}}  \\
			&&&&&&&&\\
			&&& \textbf{QMLE} & \textbf{LSE} &&&&\\
			\hline 
			\multirow{11}{*}{\textbf{Spectral}} & \multirow{3}{*}{125} & 390 & 11.122 & 11.200 & 30.419 & 9.722 & 36.540 & 36.594 \\ 
			&& 780 & 9.788 & 9.836 & 25.589 & 8.052 & 30.791 & 30.844 \\ 
			&& 2340 &7.880 & 7.907 & 22.474 & 7.413 & 26.684 & 26.594 \\ 
			&&&&&&&&\\
			& \multirow{3}{*}{250} & 390 & 8.890 & 8.951 & 29.244 & 8.622 & 37.191 & 37.164 \\ 
			&& 780 & 7.706 & 7.821 & 26.243 & 7.169 & 31.669 & 31.698 \\ 
			&& 2340 & 6.322 & 6.333 & 22.218 & 6.265 & 27.494 & 27.596 \\ 
			&&&&&&&&\\
			& \multirow{3}{*}{500} & 390 & 8.110 & 8.225 & 29.786 & 8.195 & 36.988 & 36.979 \\ 
			&& 780 & 6.233 & 6.263 & 25.712 & 6.020 & 31.907 & 31.939 \\ 
			&& 2340 & 5.186 & 5.284 & 22.122 & 5.274 & 26.590 & 26.500 \\ 
			&&&&&&&&\\ 
			&&&&&&&&\\  
			\multirow{11}{*}{\textbf{Max}} & \multirow{3}{*}{125} & 390 & 15.277 & 15.435 & 43.378 & 14.066 & 47.853 & 47.942 \\ 
			&& 780 & 13.376 & 13.534 & 37.029 & 11.495 & 40.330 & 40.321 \\ 
			&& 2340 & 11.074 & 11.226 & 30.829 & 10.276 & 33.730 & 33.581 \\ 
			&&&&&&&&\\
			& \multirow{3}{*}{250} & 390 & 12.671 & 12.755 & 41.774 & 12.871 & 47.105 & 47.194 \\ 
			&& 780 & 10.493 & 10.589 & 37.104 & 10.575 & 41.377 & 41.198 \\ 
			&& 2340 & 8.878 & 8.936 & 30.722 & 8.980 & 34.024 & 33.685 \\ 
			&&&&&&&&\\
			& \multirow{3}{*}{500} & 390 & 10.934 & 11.055 & 42.373 & 12.469 & 47.747 & 47.866 \\ 
			&& 780 & 8.478 & 8.579 & 37.289 & 9.283 & 41.139 & 40.979 \\ 
			&& 2340 & 7.360 & 7.382 & 30.253 & 8.358 & 33.399 & 33.326 \\ 
			&&&&&&&&\\ 
			&&&&&&&&\\ 
			\multirow{11}{*}{\shortstack[c] {\textbf{Relative} \\ \textbf{Frobenius}}} & \multirow{3}{*}{125} & 390 & 64.448 & 64.449 & 109.096 & 76.214 & 106.100 & 226.496 \\ 
			&& 780 & 53.078 & 53.082 & 96.772 & 76.414 & 95.346 & 190.995 \\ 
			&& 2340 & 47.542 & 47.546 & 76.710 & 66.457 & 76.359 & 147.995 \\ 
			&&&&&&&&\\
			& \multirow{3}{*}{250} & 390 & 63.974 & 63.975 & 109.525 & 76.329 & 105.938 & 226.272 \\ 
			&& 780 & 52.585 & 52.587 & 97.168 & 76.360 & 95.480 & 191.095 \\ 
			&& 2340 & 47.061 & 47.062 & 77.109 & 66.372 & 76.357 & 147.730 \\ 
			&&&&&&&&\\
			& \multirow{3}{*}{500} & 390 & 63.777 & 63.779 & 109.267 & 76.494 & 106.342 & 226.718 \\ 
			&& 780 & 52.371 & 52.372 & 97.104 & 76.279 & 95.315 & 190.935 \\ 
			&& 2340 & 46.849 & 46.852 & 76.690 & 66.615 & 76.464 & 147.686 \\ 
			\hline 
			\hline 
		\end{tabular}
		\caption{Mean matrix estimation errors in the spectral, max, and relative Frobenius norms for the conditional daily integrated volatility matrix given $n=125,250,500$ and $m=390,780,2340$.  
		\label{table: matrix estimation}}
	}
\end{table}

\subsection{An empirical study}
\label{SEC-empirical}

\begin{figure}[h]
\centering
\vspace*{-0.5cm} 
\includegraphics[width=\textwidth]{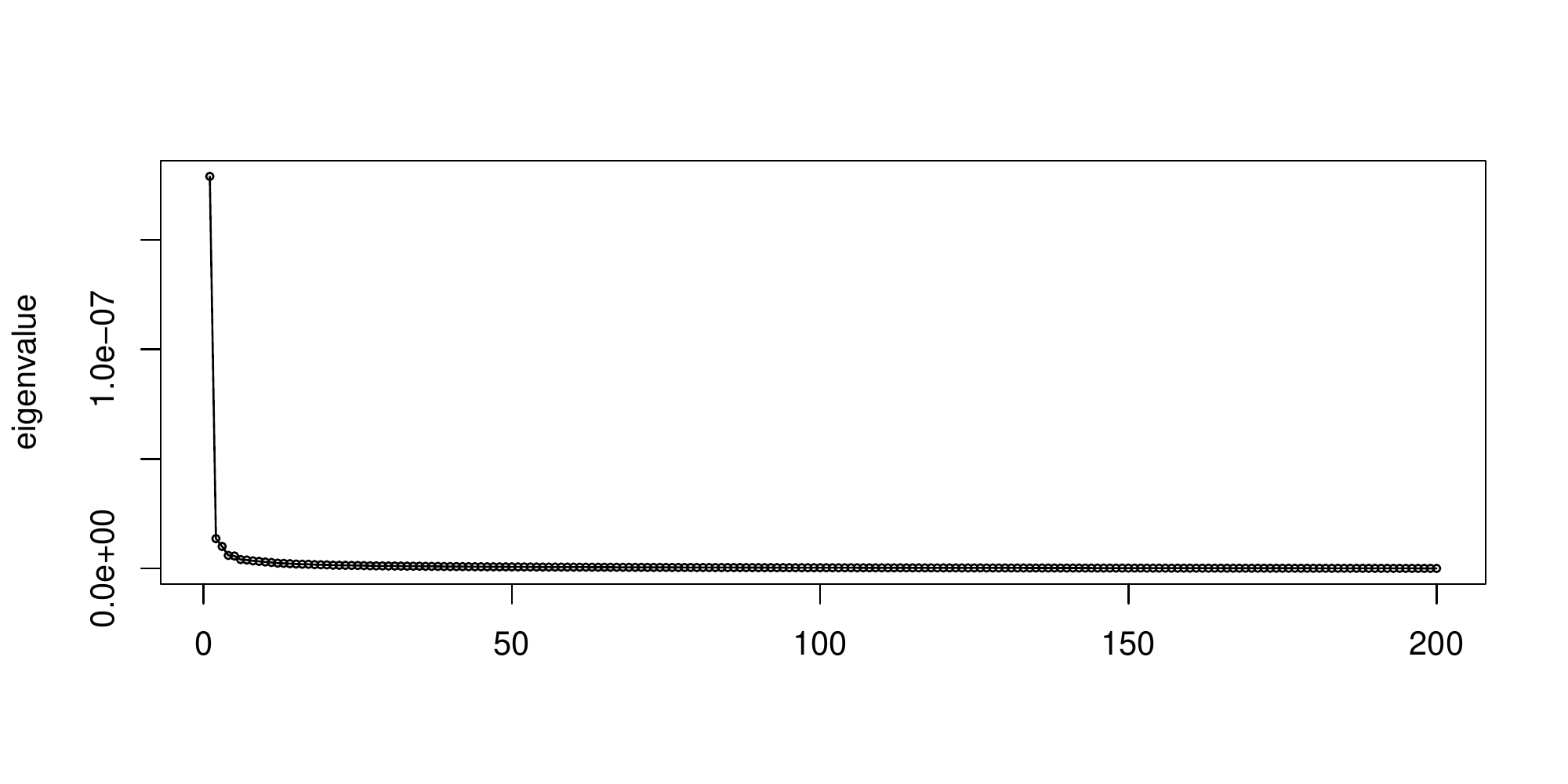} 		
\vspace*{-1.8cm} 
\caption{The eigenvalues for the sample variance of all PRVM estimators. \label{figure: eigenvalues}}
\end{figure}

\begin{figure}[h]
\vspace*{-0.5cm} 
\includegraphics[width=\textwidth]{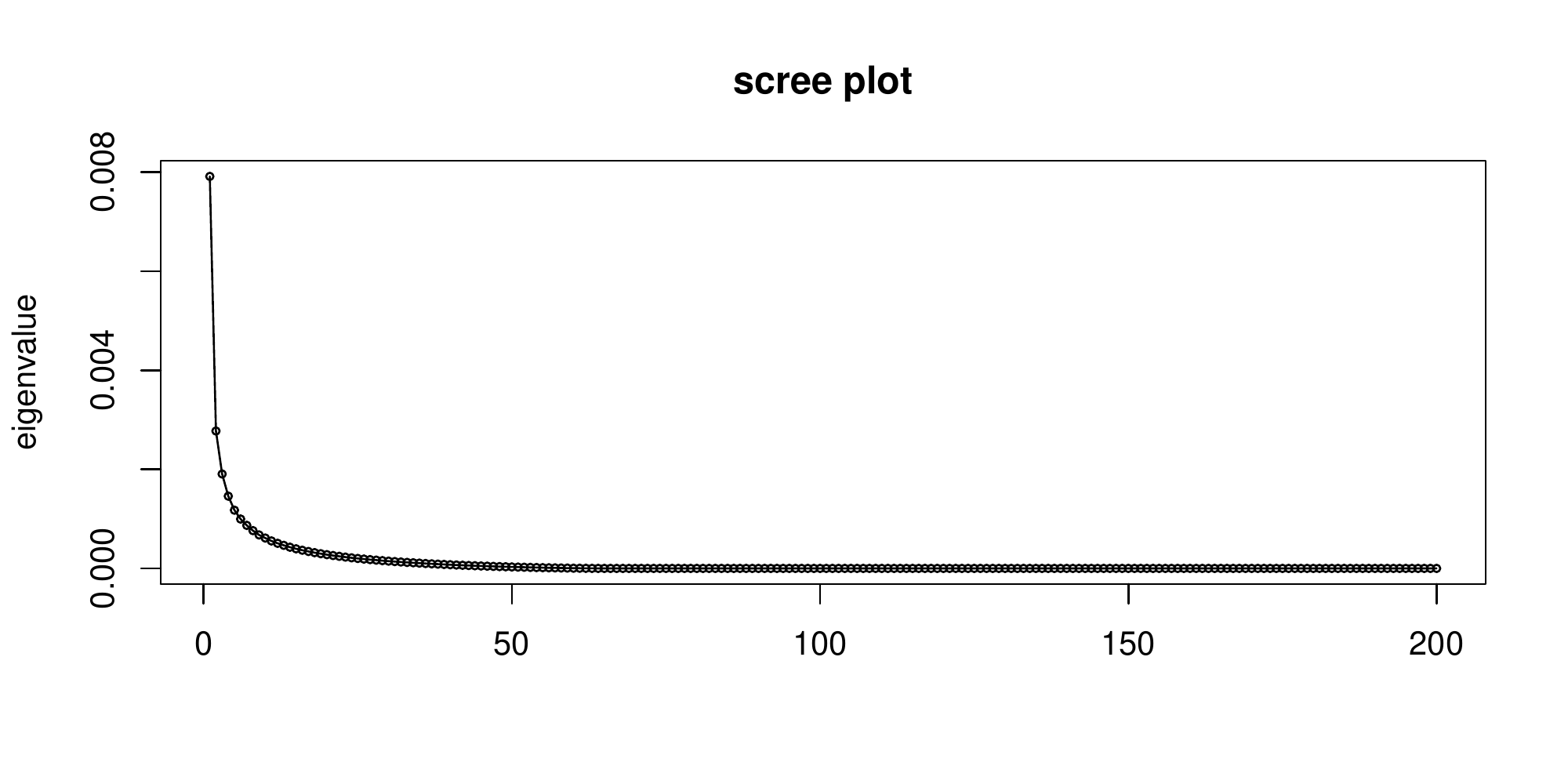}
\vspace*{-1.8cm} 
\caption{The scree plot for average eigenvalues of daily PRVM estimators. \label{figure: scree}}
\end{figure}

In this section, we demonstrate the proposed prediction methodology with real trading stock prices recorded in minute of $p=200$ companies from January 1st, 2013 to December 31st, 2013. 
The total number of low-frequency periods follows to be $n=252$ while the daily number of high-frequency returns is $m=390$. 
We estimated the daily integrated volatility matrix by the PRVM estimator \citep{christensen2010pre} and projected the obtained PRVM estimators onto the positive semi-definite cone in the spectral norm to ensure their positive semi-definiteness. 
That is, we set the negative eigenvalues to be $0$. 
The corresponding PRVM estimates are denoted as $\hbGamma_k$, $k=1, \ldots, 252$. 
The sample variance of all PRVM estimators, $\hat{\SS}_{n,m}$, was obtained and its ordered eigenvalues are presented in Figure \ref{figure: eigenvalues}. 
Moreover, let $\hlambda_{k,j}$ be the $j$th largest eigenvalue of $\hbGamma_k$, Figure \ref{figure: scree} presents the scree plot based on $\sum_{k=1}^n \hlambda_{k,1}/n$, $\sum_{k=1}^n \hlambda_{k,2}/n$, $\ldots$, $\sum_{k=1}^n \hlambda_{k,p}/n$. 
Both plots suggest that possible candidates for the number of market factors $r$ is 1,2,3,4 . 
To determine the rank $r$ specifically, we adopted the procedure as described in \cite{ait2017using} in the following:
	\begin{equation*}
	\hat{r} = \arg \underset{1 \leq j \leq r_{\max}}{\min} \sum\limits_{k=1}^{252} \[p^{-1}  \hlambda_{k,j} + j \times c_1 \left\lbrace \sqrt{\log p / m^{1/2}} + p^{-1} \log p \right\rbrace^{c_2} \] -1. 
	\end{equation*}
where we used $r_{\max}=30$, $c_1=0.02 \hlambda_{k,30}$, and $c_2=0.5$.
The procedure chose $\hat{r}=3$. 

The AR order $q$ was selected based on standard criteria such as the AIC or BIC, and the fitted model yields $q=1$. 
We optimized the proposed quasi-likelihood $\hat{L}_{n,m}(\btheta)$ and the loss function $\hat{L}^{ls}_{n,m}(\btheta)$ to obtain model parameter estimates as the follows 
	\begin{equation*}
	\hbbeta_0^{\top}=\left(1.517, -2.03, -3.918, 1.412, 9.049, -6.780 \right) \times 10^{-5}, 
	\end{equation*}
	\begin{equation*}
	\hbbeta_1= \left( 
	\begin{array}{cccccc}
	0.385 & 0.227 & 0.652 & 0.388 & 2.457 & -0.633 \\ 
	-0.001 & 0.371 & 0.128 & -0.066 & -0.097 & -0.035 \\ 
	0.014 & -0.019 & 0.224 & -0.014 & -0.382 & -0.331 \\ 
	0.022 & -0.024 & -0.063 & 0.100 & 0.100 & 0.134 \\ 
	0.023 & -0.038 & -0.052 & 0.001 & 0.437 & 0.107 \\ 
	0.015 & -0.020 & 0.009 & 0.046 & 0.236 & 0.192 \\ 
	\end{array}
	\right), 
	\end{equation*}
and 
	\begin{equation*}
	\hbbeta_0^{ls \top}=\left(1.992, -0.075, 0.132, 0.222, -0.042, 0.319 \right) \times 10^{-5}, 
	\end{equation*}
	\begin{equation*}
	\hbbeta^{ls}_1= \left(
	\begin{array}{cccccc}
	0.402 & 0.229 & 0.643 & 0.404 & 2.438 & -0.628 \\ 
	-0.005 & 0.376 & 0.128 & -0.083 & -0.106 & -0.031 \\ 
	0.023 & -0.040 & 0.225 & -0.015 & -0.384 & -0.326 \\ 
	0.023 & -0.035 & -0.055 & 0.092 & 0.097 & 0.139 \\ 
	0.005 & -0.033 & -0.057 & 0.001 & 0.445 & 0.098 \\ 
	0.005 & -0.030 & 0.015 & 0.059 & 0.232 & 0.184 \\ 
	\end{array}
	\right).
	\end{equation*}
The parameter $\bbeta_0$ denotes the intercept term in the factor volatility dynamics and its small estimated values reflect the overall level of daily factor volatilities.

To examine the model prediction performance, we carried out an out-of-sample analysis. 
In specific, we computed the proposed SV-POET estimator $\tilde{\bGamma}_k$ given observed data from low-frequency period $1$ to $k-1$. 
To obtain $\tilde{\bGamma}_k$, we first need to estimate the idiosyncratic volatility matrix $\bGamma^s$ and in the thresholding step, we used global industry classification standard (GICS) for sectors as guidance \citep{fan2015incorporating}.
Specifically, given the idiosyncratic volatility matrix estimator input $\tilde{\bGamma}^s$, we kept the volatilities within the same sector, but set the others to be zero. 
The relative prediction errors in various matrix norms: $\| \tilde{\bGamma}_k - \hbGamma_k \|_2 / \| \hbGamma_k \|_2$, $\| \tilde{\bGamma}_k - \hbGamma_k \|_F / \| \hbGamma_k \|_F$, and $\| \tilde{\bGamma}_k - \hbGamma_k \|_\text{max} / \| \hbGamma_k \|_\text{max}$ were examined.
Given any forecast origin $h$, we repeated the procedure for the remaining $n-h$ periods and obtained the mean relative prediction errors (MPEs). 
For comparison purpose, we also studied the factor GARCH-It\^o estimator and the aggregated factor GARCH-It\^o estimator for future volatility matrix $\bGamma _k$ as propoesd in \citet{kim2019factor}. 
We also used $r=3$ and employed the GICS for the thresholding step. 
For the benchmark, we as well considered the POET and PRVM methods, and predicted the future volatility matrix $\bGamma_k$ by the current volatility matrix estimators $\hbGamma^{POET}_{k-1}$ and $\hbGamma_{k-1}$. 
The GICS was employed for the thresholding step of the POET method \citep{fan2013large, fan2015incorporating}.

Table \ref{empirical: MPE} summarizes the MPE values given the SV-POET (LSE or QMLE) estimators, the  factor and aggregated factor GARCH-It\^o estimators, the POET and PRVM estimators. 
To study the dependency of model prediction performance on split points, we report the results for $h=146,168,188$ that correspond to the last trading days of July, August, September in the year 2013. 
In general, the SV-POET method outperforms the other benchmarks in predicting future volatility matrix. 
The results are consistent across various split points. 

\begin{sidewaystable}
	\centering 
	\begin{tabular}{cccccccc}
		\hline 
		\hline 
		\textbf{Matrix} & \textbf{Forecast} & \multicolumn{2}{c}{\textbf{SV-POET}} & \textbf{Factor} & \textbf{Aggregated Factor} & \multirow{2}{*}{\textbf{POET}} & \multirow{2}{*}{\textbf{PRVM}} \\
		\textbf{norms} & \textbf{origins} & QMLE & LSE & \textbf{GARCH-It\^o} & \textbf{GARCH-It\^o} && \\
		\hline 
		\multirow{3}{*}{\textbf{Spectral}} & $h=146$ & 0.853 & 0.868 & 1.163 & 0.796 & 1.055 & 1.037 \\ 
		& $h=168$ & 0.831 & 0.846 & 1.248 & 0.838 & 1.085 & 1.066 \\
		& $h=188$ & 0.811 & 0.828 & 1.233 & 0.844 & 1.045 & 1.027 \\ 
		&&&&&&& \\
		\multirow{3}{*}{\textbf{Frobenius}} & $h=146$ & 0.893 & 0.903 & 1.177 & 0.868 & 1.135 & 1.157 \\ 
		& $h=168$ & 0.882 & 0.892 & 1.229 & 0.898 & 1.157 & 1.179 \\ 
		& $h=188$ & 0.865 & 0.876 & 1.215 & 0.897 & 1.122 & 1.145 \\ 
		&&&&&&& \\
		\multirow{3}{*}{\textbf{Max}} & $h=146$ & 0.775 & 0.776 & 1.438 & 0.812 & 1.494 & 1.493 \\ 
		& $h=168$ & 0.785 & 0.787 & 1.290 & 0.811 & 1.325 & 1.324 \\ 
		& $h=188$ & 0.794 & 0.795 & 1.349 & 0.837 & 1.397 & 1.397 \\ 
		\hline 
		\hline 
	\end{tabular}
	\caption{Mean relative prediction error values for the empirical data set via the (QMLE or LSE), the factor and aggregated factor GARCH-It\^o \citep{kim2019factor}, the POET \citep{fan2013large, fan2015incorporating} and PRVM methods for forecast origin $h=146,168,188$.  \label{empirical: MPE}}
\end{sidewaystable}

We now consider the constrained portfolio allocation problem \citep{fan2012vast} with the SV-POET estimator $\tilde{\bGamma}_{k}$.  
Specifically, we minimized the following portfolio risk 
	\begin{equation*}
	\underset{\bw_k \text{ s.t. } \bw_k^{\top} \mathbf{J}=1 \text{ and } \| \bw_k \|_1=c_0 }{\text{min}} \bw_k^{\top} \tilde{\bGamma}_{k} \bw_k,
	\end{equation*}
where $\mathbf{J}=(1, \ldots, 1)^{\top} \in \RR^p$ and $c_0$ is the gross exposure constraint which varies from 1 to 2. 
The portfolio associated with $\hat{\bw}_k$ that minimizes above function is the so-called optimal portfolio. 
We computed the following out-of-sample portfolio risk for the optimal portfolio in the annualized form
	\begin{equation*}
	R_k=\sqrt{252 \cdot  \hat{\bw}_k^\top \hbGamma_k^{*} \hat{\bw}_k}
	\end{equation*}
where $\hbGamma_k^{*}$ is the realized variance of day $k$.
Given any forecast origin $h$, we repeated the procedure for the remaining $n-h$ periods and obtained the mean out-of-sample portfolio risk as 
	\begin{equation*}
	\frac{1}{n-h} \sum_{k=h+1}^n R_k. 
	\end{equation*}
For the benchmarks, we also considered the factor and aggregated factor GARCH-It\^o estimators, the POET estimator, as well as the PRVM estimator. 

\begin{figure}[h]
\centering
\includegraphics[width=\textwidth]{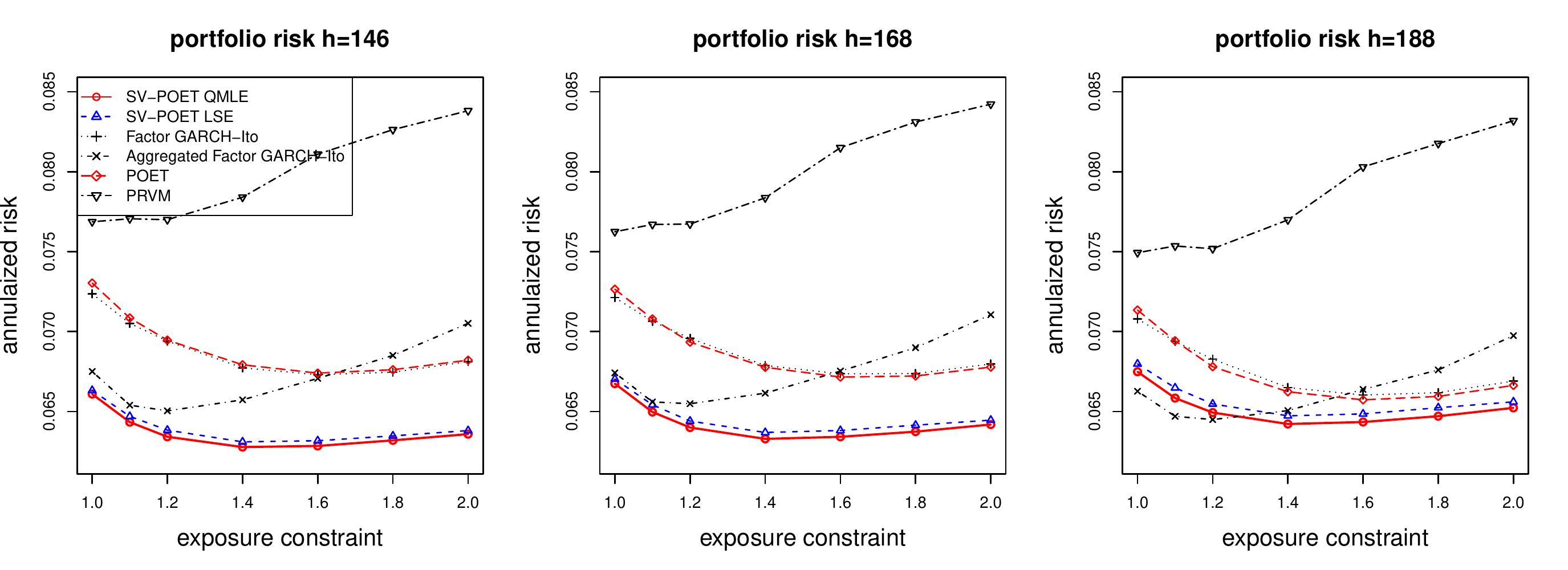} 		
\caption{Annualized mean out-of-sample portfolio risk for the optimal portfolios constructed using the SV-POET (QMLE or LSE), the factor and aggregated factor GARCH-Ito \citep{kim2019factor},  the POET \citep{fan2013large, fan2015incorporating} or the PRVM method given forecast origins $h=146,168,188$. \label{empirical-portfolio}}
\end{figure}

Figure \ref{empirical-portfolio} plots the annualized mean out-of-sample portfolio risk for various forecast origins $h=146,168,188$ and against different gross exposure constraint $c_0$ values.
The results are consistent across the split points.
The proposed SV-POET method results in smaller portfolio risk given all split points and exposure constraints comparing to the factor GARCH-It\^o, the POET and PRVM methods.
The aggregated factor GARCH-It\^o method performs well when the exposure constraint $c_0$ is small but is unstable when $c_0$ is large.
The results suggest that our proposed SV-POET model can capture the market dynamics well by utilizing the AR structure and adopting the historical integrated factor volatilities as innovations for modeling the factor volatility process.
On the other hand, the factor and aggregated factor GARCH-It\^o methods assume diagonal factor volatility that is rather restrictive, moreover, they use the GARCH type structure and adopt the squared factor returns as innovations, which may be the reasons for their suboptimal performance.
\citet{kim2019factor} further showed that the performance of their proposed methods could be improved by modeling the idiosyncratic volatility dynamics in their empirical analysis.
However, we found that the prediction performance of our proposed SV-POET method could not be enhanced by modeling the idiosyncratic volatility dynamics or by incorporating additional exogenous variables such as overnight factor returns or trading volumes when modeling the factor volatility dynamics.
These results indicate that our proposed model alone is sufficient for capturing the market dynamics, while additional information may not be very helpful.
Finally, the POET and PRVM methods do not model the volatility matrix process dynamically, which may cause a lack in their empirical performance.


\section{Conclusion} \label{SEC-6}

In this paper, we introduce a new method for vast volatility matrix estimation by employing a unified model that can accommodate both the discrete-time stochastic volatility and continuous-time It\^o diffusion models.
The proposed SV-It\^o model is capable of studying high-frequency based volatility process in the high-dimensional set-up through a low-dimensional latent factor volatility process that has an autoregressive structure.
When estimating the latent factor volatility matrices, the SV-It\^o method assumes a more general structure that is able to account for the cross-sectional market dynamics.
We note that this is important as traditional approaches that impose the diagonal assumption are rather restrictive while the corresponding models employed to study the diagonal factor volatility dynamics may not be sufficient for capturing the market dynamics.
Model parameters in the SV-It\^{o} model are estimated by either maximizing a quasi-likelihood function or minimizing a squared loss function.
The proposed LSE method is easy to implement, however, performs slightly worse than the proposed QMLE method.
However, the performance of the QMLE method depends on the initial value selection for the optimization algorithm and may not be very stable.
We show that the proposed method presents good performance in predicting future vast volatility matrix and constructing minimum-variance portfolios through our empirical study.
When comparing to other existing vast volatility estimation and prediction methods, the proposed model employs the autoregressive structure and historical factor volatilities to explain the factor volatility dynamics, which is more natural and is better supported by the empirical data.


\section*{Acknowledgements}

The research of Donggyu Kim was supported in part by KAIST Settlement/Research Subsidies for Newly-hired Faculty grant G04170049 and KAIST Basic Research Funds by Faculty (A0601003029).
The research of Xinyu Song was supported by the Fundamental Research Funds for the Central Universities (2018110128), China Scholarship Council (201806485017) and National Natural Science Foundation of China (Grant No. 11871323). 
The research of Yazhen Wang was supported in part by NSF Grants DMS-1528735, DMS-1707605, and DMS-1913149. 



\bibliography{myReferences}


\end{document}